\documentclass[10pt,twocolumn]{IEEEtran}
\usepackage{graphicx,amsmath,amsthm,amssymb,mathdots,setspace}
\newtheorem{Definition}{Definition}[section]
\newtheorem{Problem}{Problem}[section]
\newtheorem{Theorem}{Theorem}[section]
\newtheorem{Corollary}{Corollary}[section]
\newcommand{\figwidthlarge}{8.8cm}
\newcommand{\figwidthsmall}{4cm}
\newcommand{\matrixspacing}{1}
\title{Thresholds of absorbing sets in \\ Low-Density-Parity-Check codes}
\author{  Alessandro Tomasoni (alessandro.tomasoni@ieiit.cnr.it) \\ Sandro Bellini (sandro.bellini@polimi.it)\\ Marco Ferrari (marco.ferrari@ieiit.cnr.it)
\thanks{Alessandro Tomasoni and Marco Ferrari are with  Consiglio Nazionale delle Ricerche, Istituto di Elettronica e di
Ingegneria dell'Informazione
e delle Telecomunicazioni, Via G. Ponzio 34/5, 20133 Milano, Italy. Sandro Bellini is with  Politecnico di Milano, Dipartimento di Elettronica, Informazione e Bioingegneria, Piazza L. da Vinci, 20133 Milano, Italy. }}

\begin{document}
\maketitle

%%%%%%%%%%%%%%%%%%%%%%%%%%%%%%%%%%%%%%%%%%%%%%%%%%%%%%%%%%%%%%%%%%%%%%%%%%%%%%%%%%%%%%%%%%%

\begin{abstract}
In this paper, we investigate absorbing sets, responsible of error floors in Low Density Parity Check codes. We look for a concise, quantitative way to rate the absorbing sets' dangerousness. Based on a simplified model for iterative decoding evolution, we show that absorbing sets exhibit a threshold behavior. An absorbing set with at least one channel log-likelihood-ratio below the threshold can stop the convergence towards the right codeword. Otherwise convergence is guaranteed. We show that absorbing sets with negative thresholds can be deactivated simply using proper saturation levels. We propose an efficient algorithm to compute thresholds.
\end{abstract}

%%%%%%%%%%%%%%%%%%%%%%%%%%%%%%%%%%%%%%%%%%%%%%%%%%%%%%%%%%%%%%%%%%%%%%%%%%%%%%%%%%%%%%%%%%%

\section{Introduction}
\label{sec - Introduction}

 In the last years much effort has been
spent to identify the weak points of Low Density Parity Check (LDPC) code graphs, responsible for
 error floors of iterative
decoders. After the introduction of the seminal concept of
\emph{pseudocodewords}
\cite{Bib-FreKoetTIT01},\cite{Bib-KoetVonISTC03}
%and \emph{near-docodewords} \cite{McKPos}
 it is now ascertained that
these errors are caused by small subsets of nodes of the Tanner
graph that act as attractors for
 iterative decoders, even if they are
not the support of valid codewords. These structures have been
named \emph{trapping sets}
\cite{RichErrorFloors},\cite{Vasic06},\cite{ChiVasJSAC09} or
\emph{absorbing sets}
\cite{Dol_ICC07},\cite{Dol_JSAC09},\cite{Dol_TCOM09} or
\emph{absorption sets} \cite{Schlegel},
  defined in slightly different ways.
In this paper we build mainly on  \cite{Schlegel} and
\cite{Dol_ICC07},\cite{Dol_TCOM09},\cite{Dolecek}.

The first merit of \cite{Dol_ICC07},\cite{Dolecek} has been to
define Absorbing Sets (ASs) from a purely topological point of
view.
Moreover, the authors have analyzed the effects of ASs on finite
precision iterative decoders, on the basis of hardware and
Importance Sampling simulations
\cite{Dol_TCOM09},\cite{Dol_JSAC09}. ASs behavior %differently
depends on the decoder quantization and  in \cite{Dol_TCOM09} they
are classified as \emph{weak} or \emph{strong} depending on
whether they can be resolved or not by properly tuning the decoder
dynamics. In \cite{Dolecek_PostProc} the same research group
proposes a postprocessing method to resolve ASs, once the
iterative decoder is trapped.

In \cite{Schlegel} the author defines \emph{absorption sets}
(equivalent to ASs) and identifies a variety of ASs for the LDPC
code used in the IEEE 802.3an standard. The linear model of
\cite{SunTahFitz}, suitable for Min-Sum (MS) decoding, is refined
to meet the behavior of belief propagation decoders. Under some
hypotheses, the error probability level can be computed assuming
an unsaturated LDPC decoder. Loosely speaking, in this model an AS
is solved if  messages coming from the rest of the graph tend to
infinity with a rate higher than the wrong messages inside the AS.
In practical implementations, messages cannot get arbitrarily
large. Besides, hypotheses on the growth rate of the messages
entering the AS are needed. In
\cite{Schlegel} Density Evolution (DE) is used, but this is
accurate only for LDPC codes with infinite (in practice, very
large) block lengths. In \cite{Bib-ButSieALL11} the saturation is
taken into account and the input growth rate is evaluated via
Discretized DE or empirically via simulation.

In \cite{Vasic06} and successive works, the authors rate
the trapping set dangerousness with the \emph{critical number},
that is valid for hard decoders but fails to discriminate between the soft
entries of the iterative decoder.

In this paper, we look for a concise, quantitative way to rate the
ASs' dangerousness with soft decoding. We focus on Min-Sum (MS)
soft decoding that is the basis for any LDPC decoder
implementation, leaving aside more theoretical algorithms such
as Sum-Product (SPA) or Linear Programming (LP). We study the
evolution of the messages propagating inside the AS, when the
all-zero codeword is transmitted. Unlike \cite{Schlegel}, we
assume a limited dynamic of the Log Likelihood Ratios (LLRs) as in
a practical decoder implementation. The AS dangerousness can be
characterized by a \emph{threshold}  $\tau$.
We show that, under certain
hypotheses, the decoder convergence towards the right codeword
can fail only if there exist channel LLRs smaller than or equal to
$\tau$. When all channel LLRs are
 larger than $\tau$, successful decoding is assured. We also show with examples that ASs with greater $\tau$ are more harmful than ASs with smaller
$\tau$. Finally, we provide an efficient algorithm to find $\tau$.

For many ASs, $\tau<0$.
In these cases \emph{we can deactivate ASs simply setting two
saturation levels}, one for  extrinsic messages (in our system
model, this level is normalized to $1$), and another level,
smaller than $|\tau|$, for channel LLRs. This way  the code
designer can concentrate all efforts on avoiding only the
most dangerous ASs,  letting the receiver automatically deactivate
the other ones with extrinsic messages strong enough to unlock
them.

The article is organized as follows. Section \ref{sec - System
model} settles the system model. Section \ref{sec - Equilibria}
introduces the notion of equilibria and thresholds.
Section  \ref{sec - Generalized Equilibria} deals with generalized
equilibria, a tool to study ASs with arbitrary structure. Section
\ref{sec - Limit cycles} deals with limit cycles. Section \ref{sec
- Chaotic} studies the message passing behavior above threshold,
and provides a method to deactivate many ASs.
Section \ref{sec - Examples} shows
practical examples of ASs that behave as predicted by our model
during MS decoding on real complete LDPC graphs. Section \ref{sec
- Search algorithm} proposes an efficient algorithm to compute
$\tau$. Section \ref{sec - Other Properties} highlights other interesting properties.
Finally, Section  \ref{sec - Conclusions} concludes the paper.

%%%%%%%%%%%%%%%%%%%%%%%%%%%%%%%%%%%%%%%%%%%%%%%%%%%%%%%%%%%%%%%%%%%%%%%%%%%%%%%%%%%%%%%%%%%

\section{System model and definitions}
\label{sec - System model} 
\begin{figure}
\centering
  \includegraphics[width=\figwidthlarge,clip]{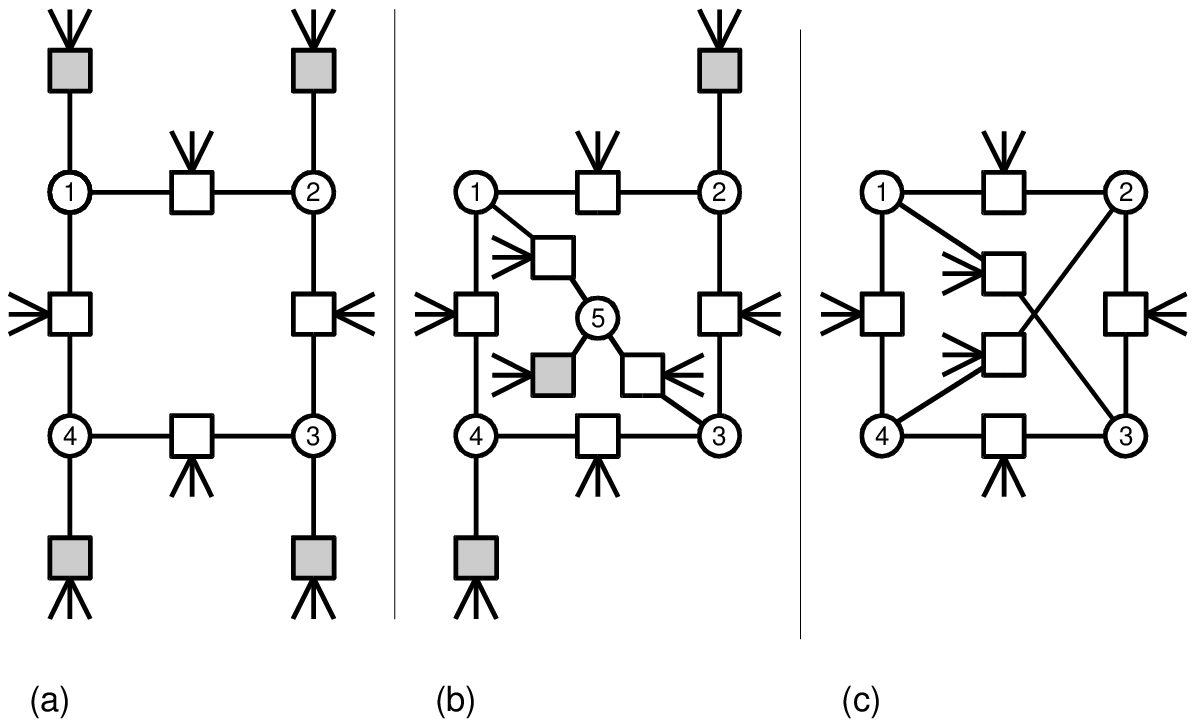}\\
 \caption{Three absorbing sets: in Fig. \ref{fig - ASs topology}(a), a maximal $(4,4)$ AS;
 in Fig. \ref{fig - ASs topology}(b), a $(5,3)$ AS; in Fig. \ref{fig - ASs topology}(c), a $(4,0)$ AS, that is also the support of a codeword.}\label{fig - ASs topology}
\end{figure}

We recall that a subset $\mathcal{D}$
of variable nodes (VNs) in a Tanner graph is an absorbing set
$(a,b)$ if \cite{Dol_ICC07}
  \begin{itemize}
    \item every VN in $\mathcal{D}$ has strictly more boundary Check Nodes (CNs) in
        $\mathcal{E}(\mathcal{D})$ than in $\mathcal{O}(\mathcal{D})$, being
       $\mathcal{E}(\mathcal{D})$  and $\mathcal{O}(\mathcal{D})$ the set of
       boundary CNs connected to $\mathcal{D}$ an \emph{even} or \emph{odd} number of
        times, respectively;
     \item the cardinality of $\mathcal{D}$ and $\mathcal{O}(\mathcal{D})$ are $a$ and $b$, respectively.
  \end{itemize}
Besides, $\mathcal{D}$ is a
\emph{fully absorbing set} if also all VNs outside $\mathcal{D}$
have strictly less neighbors in $\mathcal{O}(\mathcal{D})$ than
outside. In \cite{Dolecek} it is observed that a pattern of
all-ones for the VNs in $\mathcal{D}$ is a
  stable concurrent of the all-zeros pattern for the iterative
  bit-flipping decoder, notwithstanding a set $\mathcal{O}(\mathcal{D})$ of \emph{unsatisfied boundary
CNs} (dark CNs in Fig. \ref{fig - ASs topology}). ASs behave in a
similar manner also under iterative soft decoding, as shown and
discussed in \cite{Dol_TCOM09}, \cite{Dol_JSAC09}.

%When the rest
%of the graph has already converged towards the
%zero-codeword, the other CNs cannot detect the
%errors, since they are connected to $\mathcal{D}$ an even number
%of times. The
%decoding process can get stuck, because wrong messages are the
%majority at every VN.

If all CNs are connected to
$\mathcal{D}$ no more than twice, the AS is  \emph{elementary}. Elementary ASs are usually the most dangerous.
Given the code girth, elementary
absorbing sets can have smaller values of $a$ and $b$ than non
elementary ones \cite{Dol_ITA10}. If ASs
are the support of \emph{near-codewords} \cite{McKPos}, the
smaller $a$ is, the higher the probability of error. Besides, the
smaller the ratio $b/a$, the more dangerous the AS is
\cite{Dol_TCOM09}. In this paper we focus on elementary ASs only,
as those  in Fig. \ref{fig - ASs topology}.
An AS  is \emph{maximal} if $a=b$, as in Fig. \ref{fig - ASs topology}(a). Intuitively, maximal ASs are the mildest ones, since
they have a large number of
unsatisfied CNs. On the opposite, an AS with $b=0$ (as in Fig.
\ref{fig - ASs topology}(c)) is
the support of a codeword.

\begin{figure} \centering
  \includegraphics[width=\figwidthsmall,clip]{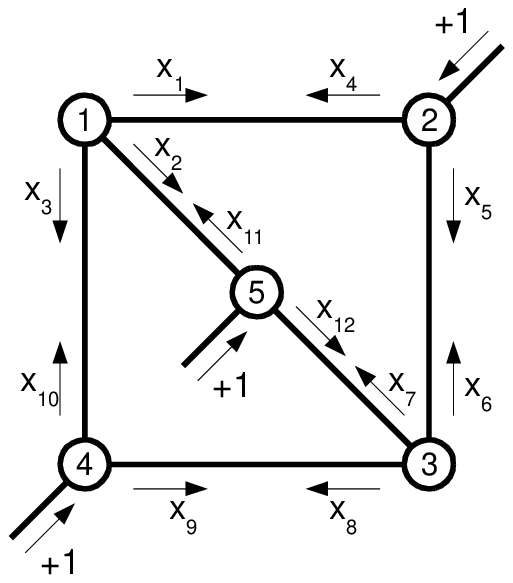}\\
  \caption{Message propagation within the AS of Fig. \ref{fig - ASs topology}(b). For simplicity, CNs in the middle of edges are not shown.}\label{fig - messages}
\end{figure}

For our analysis we assume an MS decoder, that is  insensitive to
scale factors. Thus we can normalize the maximum extrinsic message
amplitude.
We recall that, apart from saturation, the evolution of the
messages inside the AS is linear (\cite{Schlegel},
\cite{SunTahFitz}) since the CNs in $\mathcal{E}(\mathcal{D})$ simply forward the input messages.
The relation among the $N = 2 |\mathcal{E}(\mathcal{D})|$ internal
extrinsic messages $\mathbf{x}$ generated by VNs can be tracked
during the iterations, by an $N\times N$ \emph{routing} matrix
$\mathbf{A}$. Basically, $A_{i,j}=1$ iff there exists an
(oriented)
path from message $x_j$ to message $x_i$, going across one VN.
For instance, Fig. \ref{fig - messages}
depicts the LLR exchange within the AS of Fig. \ref{fig - ASs topology}(b). The corresponding first row of  $\mathbf{A}$ is\footnote{Matrix subscripts indicate subsets of rows and columns.}
\begin{equation}
  \mathbf{A}_{\{1\},\{1,\ldots,N\}}=\left[\;0\;\; 0\;\; 0\;\; 0\;\; 0\;\; 0\;\; 0\;\; 0\;\; 0\;\; 1\;\; 1\;\; 0\;\right]\,.
\end{equation}
To account for saturation we define the scalar function
$\mathrm{sat}(x) \triangleq\mathrm{sign}(x)\cdot\min\left(|x|,1\right)$
and we say that $x$ is \emph{saturated} if $|x|\geq 1$,
\emph{unsaturated} otherwise. For vectors, $\mathrm{sat}(\mathbf{x})$ is the element-wise saturation.

For the time being, we consider a \emph{parallel} message passing decoder, where all VNs are simultaneously
activated first, then all CNs are simultaneously activated, in turn\footnote{In the following, we will show that the results presented in this paper hold for any activation order of VNs and CNs, provided  extrinsic messages are propagated just once per decoding iteration.}.
The system evolution at the $k$-th iteration reads
\begin{equation}\label{eq - system evolution generic dv}
     \mathbf{x}^{(k+1)} = \mathrm{sat}\left(\mathbf{A}  \mathbf{x}^{(k)}+\mathbf{e}+\mathbf{R}\boldsymbol\lambda\right)
\end{equation}
where $\mathbf{x}^{(k)}$ are the extrinsic messages within the AS,
$    \mathbf{e}$ is the vector  of extrinsic messages entering the AS through
$\mathcal{O}(\mathcal{D})$,
and $\mathbf{R}$ is a \emph{repetition}
matrix with size $N\times a$, that constrains the $a$ channel LLRs
$\boldsymbol\lambda$ to be the same for all messages $x_j$
emanating from the same VN. Referring to Fig. \ref{fig - ASs topology}(b),
\begin{spacing}{\matrixspacing}
\begin{equation}
                     \mathbf{R}^T=\begin{bmatrix}
                                    1 & 1 & 1 & 0 & 0 & 0 & 0 & 0 & 0 & 0 & 0 & 0 \\
                                    0 & 0 & 0 & 1 & 1 & 0 & 0 & 0 & 0 & 0 & 0 & 0 \\
                                    0 & 0 & 0 & 0 & 0 & 1 & 1 & 1 & 0 & 0 & 0 & 0 \\
                                    0 & 0 & 0 & 0 & 0 & 0 & 0 & 0 & 1 & 1 & 0 & 0 \\
                                    0 & 0 & 0 & 0 & 0 & 0 & 0 & 0 & 0 & 0 & 1 & 1 \\
                                  \end{bmatrix}\,.
\end{equation}
\end{spacing}\noindent
Also note that the row weight
of $\mathbf{R}$ is unitary, i.e.
$\mathbf{R}\mathbf{1}=\mathbf{1}$.

As to the extrinsic messages
entering the AS from outside, we bypass the tricky problem of
modeling the dynamical behavior of the decoder in the whole graph
assuming that each message entering the AS has saturated to the maximal correct LLR (i.e. $+1$, since we transmit the all-zero codeword).
% is saturated to $1$
%when the rest of the graph has converged to the correct solution.
This is a reasonable hypothesis after a
sufficient number of iterations, as observed in
\cite{Dolecek_PostProc} where the authors base
their postprocessing technique on this assumption. In Section \ref{sec - Examples} we show that the decoding of a large graph is in good
agreement with the predictions of this model.
Under this hypothesis, we can  write
\begin{equation}
\mathbf{e}=\mathbf{R}\mathbf{d}-\mathbf{1}- \mathbf{A}\mathbf{1}
\end{equation}
where $\mathbf{d}$ is the $a\times 1$ vector of the VN degrees.

From now on, we will consider only left-regular LDPC codes, with
$d_i=3,\;\forall i$. Most of the theorems presented in the
following sections can be extended to a generic VN degree vector
$\mathbf{d}$. Luckily, among regular LDPC codes this is also the
case with the most favorable waterfall region. If
we set $\mathbf{d}=3\cdot \mathbf{1}$, then
$\mathbf{Rd}-\mathbf{1}=3\cdot\mathbf{R1}-\mathbf{1}=3\cdot\mathbf{1}-\mathbf{1}=2\cdot\mathbf{1}$,
and \eqref{eq - system evolution generic dv}  becomes
\begin{equation}\label{eq - system evolution dv 3}
     \mathbf{x}^{(k+1)}= \mathrm{sat}\left(\mathbf{A} \left( \mathbf{x}^{(k)}-\mathbf{1}\right)+2\cdot\mathbf{1}+\mathbf{R}\boldsymbol\lambda\right)\,.
\end{equation}
This equation is more expressive than \eqref{eq - system evolution
generic dv}, as  $\mathbf{x}^{(k)}-\mathbf{1}$ is
the gap between the current state $\mathbf{x}^{(k)}$ and the values that the
extrinsic messages should eventually achieve, once the AS is unlocked. Besides,
we will show that $-\mathbf{1}\leq\boldsymbol\lambda\leq \mathbf{1}$. Therefore, $2\cdot\mathbf{1}+\mathbf{R}\boldsymbol\lambda\geq \mathbf{1}$  and $\mathbf{A}
\left( \mathbf{x}^{(k)}-\mathbf{1}\right)\leq \mathbf{0}$.  The two competing
forces are now clearly visible. The former always helps convergence,
and the latter can amplify negative terms (if the AS is not maximal, some
rows of $\mathbf{A}$ have weight larger than $1$).

The rest of the paper is devoted to unveil the hidden properties
of \eqref{eq - system evolution dv 3}, finding sufficient
conditions for correct decoding, i.e. $\mathbf{x}^{(k)}\rightarrow\mathbf{1}$ when
$k\rightarrow\infty$.   We will assume a conservative condition
to decouple the AS behavior from the rest of the code: we do not
start with $\mathbf{x}^{(0)}=\mathbf{0}$. \emph{We take into
account any configuration of extrinsic messages $\mathbf{x}^{(0)}$ that
may result in a convergence failure}. We start
from an iteration with the rest of the decoder messages saturated
to $1$. The configuration $\mathbf{x}^{(0)}$ inside the AS, which is the result of the
message evolution up to that iteration, is unknown.
The drawback of this approach is
that we renounce to predict the probability of
message configurations inside the AS leading to decoding errors. On the
other hand, if  no $\mathbf{x}^{(0)}$   can lock the decoder, this is true
independently  of the evolution  of messages inside the AS.
We will study equilibria, limit cycles and chaotic behaviors
(i.e., aperiodic trajectories) of \eqref{eq - system evolution dv
3}, depending on the channel LLRs $\boldsymbol\lambda$ and
\emph{any} initial state $\mathbf{x}^{(0)}$.

%%%%%%%%%%%%%%%%%%%%%%%%%%%%%%%%%%%%%%%%%%%%%%%%%%%%%%%%%%%%%%%%%%%%%%%%%%%%%%%%%%%%%%%%%%%

\section{Equilibria, threshold definition and
preliminary properties} \label{sec - Equilibria}
In this Section, we study equilibria for the non-linear system
\eqref{eq - system evolution dv 3}.

\begin{Definition}\label{def - equlibrium}
A pair $(\mathbf{x},\boldsymbol\lambda)$ is  an \emph{equilibrium}  iff
 \begin{equation}
    \mathbf{x}=\mathrm{sat}\left(\mathbf{A} \left( \mathbf{x}-\mathbf{1}\right)+2\cdot\mathbf{1}+\mathbf{R}\boldsymbol\lambda\right)
\,.\end{equation}
\end{Definition}
Equilibria with $\mathbf{x}\neq\mathbf{1}$ are harmful.
They behave as attractors for the evolution of  the
extrinsic messages $\mathbf{x}^{(k)}$,
and can lead  to uncorrect
decisions. With the aim of finding the most critical ASs, those
that can lead to convergence failure even
with large values of $\boldsymbol\lambda$, we would like to solve
the following problem:
\begin{Problem}\label{prb - original problem}
    \begin{align}
    \nonumber    \tau'=\max_{\boldsymbol \lambda,\mathbf{x}}&\quad\min(\boldsymbol\lambda)\\
        s.t.
     & \quad -\mathbf{1}\leq\mathbf{x}\leq \mathbf{1},\quad
  \label{eq - x_j neq 1}    \exists j: x_j< 1
    \\
  \nonumber                  & \quad \mathbf{x}=\mathrm{sat}\left(\mathbf{A} \left( \mathbf{x}-\mathbf{1}\right)+2\cdot\mathbf{1}+\mathbf{R}\boldsymbol\lambda\right)
 \,.\end{align}
\end{Problem}
The  constraint \eqref{eq - x_j neq 1}  restricts the search to
bad equilibria, having at least one extrinsic message smaller than
$1$. We call $\tau'$ the
\emph{threshold}, since the AS has no
bad equilibria with $\boldsymbol\lambda$ above that value. In Section \ref{sec - Chaotic}
we will show that the notion of threshold does not pertain only to
bad equilibria, but also to any other bad trajectory of \eqref{eq
- system evolution dv 3}, not achieving
$\mathbf{x}^{(\infty)}=\mathbf{1}$.

In the above optimization problem, for simplicity we did not
assign upper and lower bounds to the channel LLRs
$\boldsymbol\lambda$. In practice,
we can restrict our search in the range
$-\mathbf{1}\leq\boldsymbol\lambda\leq\mathbf{1}$.
\begin{Theorem}\label{thm
- equilibrium -1} The pair
$(\mathbf{x}=-\mathbf{1},\boldsymbol\lambda=-\mathbf{1})$ is
always an equilibrium.
\end{Theorem}
\begin{IEEEproof} Substituting
$(\mathbf{x}=-\mathbf{1},\boldsymbol\lambda=-\mathbf{1})$ in the equilibrium
equation, we obtain
\begin{multline}
\mathrm{sat}\left(\mathbf{A} \left( \mathbf{x}-\mathbf{1}\right)+2\cdot\mathbf{1}+\mathbf{R}\boldsymbol\lambda'\right)
=\mathrm{sat}\left(-2\cdot\underbrace{\mathbf{A}\mathbf{1}}_{\geq \mathbf{1}}+\mathbf{1}\right)\\=-\mathbf{1}=\mathbf{x}\,.
\end{multline}
\end{IEEEproof}
\begin{Theorem}\label{thm - equilibrium +1}
The only equilibrium $(\mathbf{x},\boldsymbol\lambda)$ for a system having
$\boldsymbol\lambda>\mathbf{1}$ and
$-\mathbf{1}\leq\mathbf{x}\leq\mathbf{1}$ is in $\mathbf{x}=\mathbf{1}$.
\end{Theorem}
\begin{IEEEproof} Since  $\boldsymbol\lambda>\mathbf{1}$, we can define a
strictly positive quantity $\Delta\triangleq\min(\boldsymbol\lambda)-1>0$.
Consider parallel message passing. Focusing on the evolution of \eqref{eq - system evolution dv 3},
  \begin{multline}
      \mathbf{x}^{(1)}=\mathrm{sat}\left(\mathbf{A}\underbrace{(\mathbf{x}^{(0)}-\mathbf{1})}_{\geq-2\cdot\mathbf{1}}+\underbrace{2\cdot \mathbf{1}+\mathbf{R}\boldsymbol\lambda}_{\geq(3+\Delta)\mathbf{1}}\right)\\\geq\mathrm{sat}\left(-2\underbrace{\mathbf{A}\mathbf{1}}_{\leq2\cdot\mathbf{1}}+(3+\Delta)\mathbf{1}\right)
      \geq\mathrm{sat}(-1+\Delta)\cdot\mathbf{1}\,.
    \end{multline}
    If $\mathrm{sat}(-1+\Delta)=1$, then $\mathbf{x}^{(1)}=\mathbf{1}$, and we stop. Otherwise,  $\mathrm{sat}(-1+\Delta)=-1+\Delta$ and we go on. For the generic step, assuming $\mathbf{x}^{(k)}\geq \mathrm{sat}\left(-1+k\Delta\right)\cdot\mathbf{1}$ and proceeding by recursion, we have
\begin{multline}
      \mathbf{x}^{(k+1)}=\mathrm{sat}\left(\mathbf{A}\underbrace{(\mathbf{x}^{(k)}-\mathbf{1})}_{\geq(-2+k\Delta)\cdot\mathbf{1}}+\underbrace{2\cdot \mathbf{1}+\mathbf{R}\boldsymbol\lambda}_{\geq(3+\Delta)\mathbf{1}}\right)\\\geq\mathrm{sat}\left(-\mathbf{1}+k\Delta\underbrace{\mathbf{A}\mathbf{1}}_{\geq \mathbf{1}}+\Delta\mathbf{1}\right)
  \geq\mathrm{sat}\left(-1+(k+1)\Delta\right)\cdot\mathbf{1}\,.
    \end{multline}
The same inequalities hold also in case of sequential message passing, activating CNs in arbitrary order (once per iteration).

As soon as $-1+(k+1)\Delta\geq 1$, the recursion ends. We conclude that the message passing algorithm will eventually achieve  $\mathbf{x}=\mathbf{1}$. \end{IEEEproof}

Being $\tau'$ the result of
a maximization, a straight consequence of the above two theorems is
\begin{Corollary} \label{thm - alpha range}
As for Problem \ref{prb - original problem}, $-1\leq\tau'\leq 1$.
\end{Corollary}
The two boundary values $\tau'=-1$ and $\tau'=1$ are the thresholds of maximal
ASs and codewords, respectively:
\begin{Theorem}
Any support of a codeword has $\tau'=1$.
Maximal absorbing sets have $\tau'=-1$.
\end{Theorem}
\begin{IEEEproof} We start from codewords.
$(\mathbf{x}=-\mathbf{1},\boldsymbol\lambda=\mathbf{1})$ is a valid
equilibrium for Problem \ref{prb - original problem}. Indeed:
\begin{multline}
  \mathrm{sat}\left(\mathbf{A}(\mathbf{x}-\mathbf{1})+2\cdot\mathbf{1}+\mathbf{R}\boldsymbol\lambda\right)=\mathrm{sat}\left(-2\underbrace{\mathbf{A}\mathbf{1}}_{=2\cdot\mathbf{1}}+3\cdot\mathbf{1}\right)\\=\mathrm{sat}\left(-4\cdot\mathbf{1}+3\cdot\mathbf{1}\right)=-\mathbf{1}=\mathbf{x}\,.
\end{multline}
By Corollary \ref{thm - alpha range}  we conclude that $\tau'=1$.

Referring to maximal ASs, for any
$-\mathbf{1}\leq\mathbf{x}^{(0)}\leq\mathbf{1}$ and
$\boldsymbol\lambda>-\mathbf{1}$,
we can
define a strictly positive quantity
$\Delta\triangleq\min(\boldsymbol\lambda)-(-1)>0$. Focusing on the evolution
of \eqref{eq - system evolution dv 3},
    \begin{multline}
      \mathbf{x}^{(1)}=\mathrm{sat}\left(\mathbf{A}\underbrace{(\mathbf{x}^{(0)}-\mathbf{1})}_{\geq-2\cdot\mathbf{1}}+\underbrace{2\cdot \mathbf{1}+\mathbf{R}\boldsymbol\lambda}_{\geq(1+\Delta)\mathbf{1}}\right)\\\geq\mathrm{sat}\left(-2\underbrace{\mathbf{A}\mathbf{1}}_{=\mathbf{1}}+(1+\Delta)\mathbf{1}\right)
      \geq\mathrm{sat}(-1+\Delta)\cdot\mathbf{1}\,.
    \end{multline}
If $\mathrm{sat}(-1+\Delta)=1$, then $\mathbf{x}^{(1)}=\mathbf{1}$, and we
stop. Otherwise,  $\mathrm{sat}(-1+\Delta)=-1+\Delta$ and we go on. For the
generic step, assuming $\mathbf{x}^{(k)}\geq
\mathrm{sat}\left(-1+k\Delta\right)\cdot\mathbf{1}$ and proceeding by
recursion, we obtain
   \begin{multline}
      \mathbf{x}^{(k+1)}=\mathrm{sat}\left(\mathbf{A}\underbrace{(\mathbf{x}^{(k)}-\mathbf{1})}_{\geq(-2+k\Delta)\cdot\mathbf{1}}+\underbrace{2\cdot \mathbf{1}+\mathbf{R}\boldsymbol\lambda}_{\geq(1+\Delta)\mathbf{1}}\right)\\\geq \mathrm{sat}\left(-1+(k+1)\Delta\right)\cdot\mathbf{1}\,.
    \end{multline}
As soon as $-1+(k+1)\Delta\geq 1$, the recursion ends.
The message passing algorithm will eventually achieve  $\mathbf{x}=\mathbf{1}$, that is not a valid equilibrium
for Problem \ref{prb - original problem}. We conclude that at least one
element in $\boldsymbol\lambda$ must be equal to $-1$, therefore $\tau'=-1$.
\end{IEEEproof}

%%%%%%%%%%%%%%%%%%%%%%%%%%%%%%%%%%%%%%%%%%%%%%%%%%%%%%%%%%%%%%%%%%%%%%%%%%%%%%%%%%%%%%%%%%%

\section{Generalized equilibria} \label{sec - Generalized Equilibria}
Most of the effort of this Section is in the
reformulation of Problem  \ref{prb - original problem}, to make it
manageable. First, in place of equilibria, we consider a slightly more general
case, removing the repetition matrix $\mathbf{R}$ and assuming $N$ unconstrained channel LLRs  $\boldsymbol\lambda$.
\begin{Definition}\label{def - generalized equilibrium}
A pair $(\mathbf{x},\boldsymbol\lambda)$ is  a \emph{generalized equilibrium}  iff
 \begin{equation}\label{eq - generalized equilibrium}
 \mathbf{x}=\mathrm{sat}\left(\mathbf{A} \left( \mathbf{x}-\mathbf{1}\right)+2\cdot\mathbf{1}+\boldsymbol\lambda\right)\,.
 \end{equation}
\end{Definition}
Accordingly, we write the following optimization problem.
\begin{Problem}\label{prb - generalized problem}
    \begin{align}
    \nonumber  \tau^*= \max_{\boldsymbol \lambda,\mathbf{x}}&\quad\min(\boldsymbol\lambda)\\
     \nonumber   s.t.
     & \quad -\mathbf{1}\leq\mathbf{x}\leq \mathbf{1}, \quad \exists j: x_j< 1
    \\
  \nonumber                  & \quad \mathbf{x}=\mathrm{sat}\left(\mathbf{A} \left( \mathbf{x}-\mathbf{1}\right)+2\cdot\mathbf{1}+\boldsymbol\lambda\right).
 \end{align}
\end{Problem}

The following theorem  holds.
\begin{Theorem}\label{thm - tau'=tau*}
As for Problems \ref{prb - original problem} and \ref{prb -
generalized problem}, $\tau'=\tau^*$.
\end{Theorem}
\begin{IEEEproof} We show that $\tau'\leq\tau^*$ and  $\tau'\geq\tau^*$.

Every equilibrium is also a generalized equilibrium. Given a solution $(\mathbf{x}',\boldsymbol\lambda')$ of Problem \ref{prb -
original problem} with $\min(\boldsymbol\lambda')=\tau'$,
the solution $(\mathbf{x}^*,\boldsymbol\lambda^*)$ with
$\mathbf{x}^*=\mathbf{x}'$ and
$\boldsymbol\lambda^*=\mathbf{R}\boldsymbol\lambda'$ satisfies the constraints of Problem \ref{prb -
original problem}.
 Being
$\tau^*$ the result of the maximization in Problem \ref{prb - generalized
problem}, we conclude that $\tau^*\geq \tau'$.

  On the converse,  generalized equilibria may not be equilibria. Indeed,
$\boldsymbol\lambda^*$ could not be compatible with the
repetition forced by matrix $\mathbf{R}$.
Notwithstanding this, if a generalized equilibrium $( \mathbf{x}^*, \boldsymbol\lambda^*)$ exists,
then also an equilibrium $(\mathbf{x}', \boldsymbol\lambda')$ exists, with
$\mathbf{x}'\leq \mathbf{x}^*$ and
$\boldsymbol\lambda'=\min(\boldsymbol\lambda^*)\cdot \mathbf{1}$.
 Consider channel LLRs
$\boldsymbol\lambda'=\min(\boldsymbol\lambda^*)\cdot \mathbf{1}$. We
explicitly provide an initialization $\mathbf{x}^{(0)}$ for \eqref{eq - system
evolution dv 3} that makes the extrinsic messages achieve an equilibrium
$(\mathbf{x}', \boldsymbol\lambda')$, with $\mathbf{x}'\leq \mathbf{x}^*$.
First, note that
\begin{multline}
     \mathbf{x}^{(k+1)}= \mathrm{sat}\left(\mathbf{A} \left( \mathbf{x}^{(k)}-\mathbf{1}\right)+2\cdot\mathbf{1}+\mathbf{R}\boldsymbol\lambda'\right)\\=\mathrm{sat}\left(\mathbf{A} \left( \mathbf{x}^{(k)}-\mathbf{1}\right)+2\cdot\mathbf{1}+\min(\boldsymbol\lambda^*)\underbrace{\mathbf{R}\mathbf{1}}_{=\mathbf{1}}\right)\,.
\end{multline}
If we set $\mathbf{x}^{(0)}=\mathbf{x}^*$, we obtain the inequality
\begin{multline}
    \mathbf{x}^{(1)}=\mathrm{sat}\left(\mathbf{A} \left( \mathbf{x}^*-\mathbf{1}\right)+2\cdot\mathbf{1}+\min(\boldsymbol\lambda^*)\mathbf{1}\right)\\\leq\mathrm{sat}\left(\mathbf{A} \left( \mathbf{x}^*-\mathbf{1}\right)+2\cdot\mathbf{1}+\boldsymbol\lambda^*\right)=\mathbf{x}^*=\mathbf{x}^{(0)}\,.
\end{multline}
Proceeding by induction,
\begin{multline}
     \mathbf{x}^{(k+1)}= \mathrm{sat}\left(\mathbf{A} \left( \mathbf{x}^{(k)}-\mathbf{1}\right)+2\cdot\mathbf{1}+\min(\boldsymbol\lambda^*)\mathbf{1}\right)\\   \leq \mathrm{sat}\left(\mathbf{A} \left( \mathbf{x}^{(k-1)}-\mathbf{1}\right)+2\cdot\mathbf{1}+\min(\boldsymbol\lambda^*)\mathbf{1}\right)=\mathbf{x}^{(k)}
\end{multline}
since $A_{i,j}\geq 0,\;\forall i,j$. The above equation states that the sequence
$\{\mathbf{x}^{(k)}\}$ is monotonically decreasing. Yet, it cannot assume
arbitrarily small values, since extrinsic messages have a lower saturation to
$-1$. We conclude that $\{\mathbf{x}^{(k)}\}$ must achieve a new equilibrium
$\mathbf{x}'\leq \mathbf{x}^*$.

The equilibrium $(\mathbf{x}',\boldsymbol\lambda')$ satisfies all the constraints of Problem
\ref{prb - original problem}. Being $\tau'$ the result of a maximization, $\tau'\geq \tau^*$. \end{IEEEproof}

The above statements do not claim that the two problems are
equivalent. Indeed, they can be maximized by \emph{different} pairs
$(\mathbf{x},\boldsymbol\lambda)$. Anyway, as long as we are interested in the
AS threshold, we can deal with Problem \ref{prb - generalized problem} instead
of Problem \ref{prb - original problem} and with generalized equilibria
instead of equilibria.
%%%%%%%%%%%%%%%%%%%%%%%%%%%%%%%%%%%%%%%%%%%%%%%%%%%%%%%%%%%%%%%%%%%%%%%%%%%%%%%%%%%%%%%%%%%

\section{Limit cycles}\label{sec - Limit cycles}
In this Section, we focus on limit cycles, i.e. on extrinsic messages that
periodically take the same values. We show that they have thresholds smaller than or equal to equilibria. Therefore, we will neglect them.
\begin{Definition}\label{def - limit cycle}
The sequence
$(\{\mathbf{x}''^{(0)},\ldots,\mathbf{x}''^{(L-1)}\},\boldsymbol\lambda'')$
is a \emph{limit cycle} with period $L$ iff $\forall k$,
\begin{equation}
  \mathbf{x}''^{(k+1\:\mathrm{mod}\:L)}=\mathrm{sat}\left(\mathbf{A} \left( \mathbf{x}''^{(k \:\mathrm{mod}\: L)}-\mathbf{1}\right)+2\cdot\mathbf{1}+\mathbf{R}\boldsymbol\lambda''\right)\,.
\end{equation}
\end{Definition}
Limit cycles can be interpreted as equilibria of the \emph{augmented} AS,
described by an augmented matrix $\mathbf{A}''$ of size
$(NL)\times(NL)$. While the VN and CN activation order does not matter in case
of equilibria  (at the equilibrium, extrinsic
messages do not change if we update them all together, or one by one in
arbitrary order), this is not true in case of limit cycles. Indeed, the
associated set of equations depends on the  decoding order.

In case of parallel message passing, one can
write a system of equations with $NL$ rows, where the $l$-th horizontal stripe
of $N$ equations represents the evolution of extrinsic messages from state
$\mathbf{x}''^{(l-1 \mod L)}$ to  $\mathbf{x}''^{(l)}$
\begin{align}\label{eq - augmented system}
    \begin{bmatrix}
      \mathbf{x}''^{(0)} \\
      \mathbf{x}''^{(1)} \\
      \vdots\\
      \mathbf{x}''^{(L-1)} \\
    \end{bmatrix}
=\mathrm{sat}\left(\mathbf{A}''
\left(
    \begin{bmatrix}
      \mathbf{x}''^{(0)} \\
      \mathbf{x}''^{(1)} \\
      \vdots\\
      \mathbf{x}''^{(L-1)} \\
    \end{bmatrix}
    -\mathbf{1}\right)+
     2\cdot
    \mathbf{1}
    +
    \begin{bmatrix}
      \mathbf{R}\boldsymbol\lambda'' \\
       \mathbf{R}\boldsymbol\lambda''  \\
      \vdots\\
      \mathbf{R}\boldsymbol\lambda'' \\
    \end{bmatrix}
   \right)\,.
\end{align}

Instead, in sequential (or serial-C \cite{ShaLit}) decoding  CNs
are activated one by one, in turn, immediately updating the
a-posteriori LLRs of the VNs connected thereto. The augmented
matrix changes, since only the first CNs use extrinsic messages
produced at the previous iteration, while all others exploit
messages generated during the same iteration. We can represent
this behavior defining two matrices $\mathbf{\bar{A}}$ and
$\mathbf{\underline{A}}$, binary partitions of $\mathbf{A}$,
\begin{equation}
  \bar{A}_{i,j}, \underline{A}_{i,j} \in \{0,1\},\qquad
  \mathbf{\bar{A}}+  \mathbf{\underline{A}} = \mathbf{A}
\end{equation}
and writing an  augmented matrix as
\begin{spacing}{\matrixspacing}
\begin{equation}\label{eq - Aaugmented}
    \mathbf{A}''=
\begin{bmatrix}
     \mathbf{\underline{A}}& \mathbf{0} &  \cdots & \mathbf{0} & \mathbf{\bar{A}}\\
  \mathbf{\bar{A}} &   \mathbf{\underline{A}}& \cdots & \mathbf{0} & \mathbf{0} \\
   \mathbf{0} & \mathbf{\bar{A}} &    & \mathbf{0} &  \mathbf{0}\\
   \vdots & \vdots& \ddots & \ddots & \vdots \\
  \mathbf{0} & \mathbf{0} &  \cdots & \mathbf{\bar{A}} &  \mathbf{\underline{A}} \\
\end{bmatrix}\,.
\end{equation}
\end{spacing}
$\bar{\mathbf{A}}$ and $\underline{\mathbf{A}}$ have upper and lower triangular shapes, due to the sequential update order.
Note that \eqref{eq - Aaugmented} is valid not
only for sequential CN message passing decoding, but also for any arbitrary
order\footnote{In this case, the lower and upper triangular shape is lost.}, as long as all extrinsic messages are activated in
turn, once per decoding iteration. Parallel message passing is a special case of \eqref{eq - Aaugmented}, with
$\mathbf{\bar{A}}=\mathbf{A}$ and $\mathbf{\underline{A}}=\mathbf{0}$.
Therefore we provide the following theorem only for the most
general case.
\begin{Theorem}\label{thm - sequential cycles do not matter}
  If there exists a limit cycle  $(\{\mathbf{x}''^{(0)},\mathbf{x}''^{(1)},\ldots,\mathbf{x}''^{(L-1)}\},\boldsymbol\lambda'')$,   a generalized equilibrium $(\mathbf{x}^*,\boldsymbol\lambda^*)$ with $\boldsymbol\lambda^*\geq\mathbf{R}\boldsymbol\lambda''$ exists, too.
\end{Theorem}
\begin{IEEEproof} Consider any partition of the identity matrix $\mathbf{I}$ in
$L$ binary matrices, with size $N\times N$:
\begin{equation}
  W_{i,j}^{(l)} \in \{0,1\},\qquad
  \sum_{l=0}^{L-1}\mathbf{W}^{(l)}= \mathbf{I}\,.
\end{equation}
Then
\begin{align}
\nonumber   \sum_{l=0}^{L-1}\mathbf{W}^{(l)}\mathbf{x}''^{(l)} =& \; \sum_{l=0}^{L-1}\mathbf{W}^{(l)}\mathrm{sat}\Big(\mathbf{\bar{A}}(\mathbf{x}''^{(l-1 \mod L)}-\mathbf{1})\\\nonumber&+  \mathbf{\underline{A}}(\mathbf{x}''^{(l)}-\mathbf{1})+2\cdot\mathbf{1}+\mathbf{R}\boldsymbol\lambda''\Big) \\\nonumber
   =&\; \mathrm{sat}\Bigg(\sum_{l=0}^{L-1}\mathbf{W}^{(l)}\mathbf{\bar{A}}(\mathbf{x}''^{(l-1 \mod L)}-\mathbf{1})\\&+\sum_{l=0}^{L-1}\mathbf{W}^{(l)}  \mathbf{\underline{A}}(\mathbf{x}''^{(l)}-\mathbf{1})
   \\
  \nonumber &+2\cdot\underbrace{\sum_{l=0}^{L-1}\mathbf{W}^{(l)}}_{=\mathbf{I}}\mathbf{1}+\underbrace{\sum_{l=0}^{L-1}\mathbf{W}^{(l)}}_{=\mathbf{I}}\mathbf{R}\boldsymbol\lambda''\Bigg)
\end{align}
where in the second equation  $\mathbf{W}^{(l)}$ enters into the
$\mathrm{sat}(\cdot)$ function since
$1\cdot\mathrm{sat}(x)=\mathrm{sat}(1\cdot x)$ and
$0\cdot\mathrm{sat}(x)=\mathrm{sat}(0\cdot x),\;\forall x$.

Choose a vector $\mathbf{x}^*$ of extrinsic messages as
\begin{equation}
   x_j^*=\min_l\left({x''_j}^{(l)}\right),\;\forall j\,.
\end{equation}
As a consequence, we have  $\mathbf{x}^*\leq \mathbf{x}''^{(l)},\;\forall l$ and
 \begin{align}
\nonumber   \sum_{l=0}^{L-1}\mathbf{W}^{(l)}\mathbf{x}''^{(l)} =&\; \mathrm{sat}\left(\underbrace{\sum_{l=0}^{L-1}\mathbf{W}^{(l)}}_{=\mathbf{I}}\mathbf{\bar{A}}(\mathbf{x}^*-\mathbf{1})\right.\\
\nonumber&\left.+\underbrace{\sum_{l=0}^{L-1}\mathbf{W}^{(l)}}_{=\mathbf{I}}  \mathbf{\underline{A}}(\mathbf{x}^*-\mathbf{1})+2\cdot\mathbf{1}+\underbrace{\mathbf{R}\boldsymbol\lambda''+\mathbf{\Delta}}_{\triangleq\boldsymbol\lambda^*}\right) \\
 =&\; \mathrm{sat}\left(\mathbf{A}(\mathbf{x}^*-\mathbf{1})+2\cdot\mathbf{1}+\boldsymbol\lambda^*\right)
\end{align}
with $\mathbf{\Delta}=\sum_{l=0}^{L-1}\mathbf{W}^{(l)}\mathbf{\bar{A}}(\mathbf{x}''^{(l-1 \mod L)}-\mathbf{x}^*)+\sum_{l=0}^{L-1}\mathbf{W}^{(l)}  \mathbf{\underline{A}}(\mathbf{x}''^{(l)}-\mathbf{x}^*)\geq 0$, being    $A_{i,j}\geq0,\forall i,j$.
Finally, choose the partition $\left\{\mathbf{W}^{(l)}\right\}$ that
implements the $\min(\cdot)$ function, i.e.,
\begin{equation}
     \sum_{l=0}^{L-1}\mathbf{W}^{(l)}\mathbf{x}''^{(l)} = \mathbf{x}^*
\end{equation}
thus achieving a generalized equilibrium $\left(\mathbf{x}^*,\boldsymbol\lambda^*\right)$  with $\boldsymbol\lambda^*= \mathbf{R}\boldsymbol\lambda''+\mathbf{\Delta}\geq  \mathbf{R}\boldsymbol\lambda''$.
\end{IEEEproof}

A straight consequence of Theorem \ref{thm - sequential cycles do not matter}
is that limit cycles can be neglected, when we compute the AS threshold.

%%%%%%%%%%%%%%%%%%%%%%%%%%%%%%%%%%%%%%%%%%%%%%%%%%%%%%%%%%%%%%%%%%%%%%%%%%%%%%%%%%%%%%%%%%%

\section{Behavior analysis above threshold}\label{sec - Chaotic}
We also have to take into account potential chaotic behaviors of the
extrinsic messages in \eqref{eq - system evolution dv 3}. In principle,
$\mathbf{x}^{(k)}$ could even evolve without achieving any equilibrium or
limit cycle.
Yet, above the
threshold $\tau$ the extrinsic messages $\mathbf{x}^{(k)}$ achieve
$\mathbf{1}$.
\begin{Theorem}\label{thm - chaotic behaviors}
Let $\tau$ be the solution of Problem
\ref{prb - original problem} or \ref{prb - generalized problem}.  If $\boldsymbol\lambda> \tau\cdot \mathbf{1}$,
for any starting $\mathbf{x}^{(0)}$ with $-\mathbf{1}\leq
\mathbf{x}^{(0)} \leq \mathbf{1}$, for a sufficiently large $K\geq
0$
\begin{equation}
    \mathbf{x}^{(k)}=\mathbf{1},\quad\forall k\geq K\,.
\end{equation}
\end{Theorem}
\begin{IEEEproof} For the time being, consider channel messages
$\mathbf{\hat{\boldsymbol\lambda}}$ that can assume only quantized values
between $-1$ and $1$, with uniform step $\delta=\frac{1}{Q}$, $Q\in
\mathbb{N}$. Assume that also extrinsic messages $\mathbf{\hat{x}}$ are
quantized numbers, with the same step $\delta$. Therefore,
$\mathbf{\hat{x}}^{(0)}$ can only assume $(2Q)^N$ different values. Letting
the system
\begin{equation}
     \mathbf{\hat{x}}^{(k+1)}= \mathrm{sat}\left(\mathbf{A} \left( \mathbf{\hat{x}}^{(k)}-\mathbf{1}\right)+2\cdot\mathbf{1}+\mathbf{R}\mathbf{\hat{\boldsymbol\lambda}}\right)
\end{equation}
evolve, it is clear that  extrinsic messages at every time  $k>0$ must belong
to the same set of $(2Q)^N$  values. When
$\mathbf{\hat{\boldsymbol\lambda}}>\tau\cdot\mathbf{1}$, the analysis
presented in previous Sections assures that the only remaining equilibrium is
$\mathbf{x}=\mathbf{1}$. Indeed, other equilibria cannot exist since they
would need $\min\left(\hat{\boldsymbol\lambda}\right)\leq\tau$. By Theorem  \ref{thm - sequential cycles do not matter}, also limit cycles do
not exist, both in case of parallel and sequential decoding. Therefore, the
only value that $\mathbf{\hat{x}}$ can assume more than once is $\mathbf{1}$
(otherwise we would incur in equilibria or cycles). We can conclude that
$\mathbf{x}=\mathbf{1}$ will be reached in at most $K=(2Q)^N$ iterations.
After that, the extrinsic messages will remain constant and the absorbing set
will be defused.

If $\mathbf{x}^{(0)}$ and $\mathbf{\boldsymbol\lambda}$ are not quantized, we
can always identify a sufficiently small quantization step $\delta$ and a
quantized pair
$\left(\mathbf{\hat{x}}^{(0)},\mathbf{\hat{\boldsymbol\lambda}}\right)$
s.t.
\begin{equation}
    \mathbf{\hat{x}}^{(0)}\leq \mathbf{x}^{(0)},\qquad
    \tau\cdot \mathbf{1}<\mathbf{\hat{\boldsymbol\lambda}}\leq \boldsymbol\lambda
\end{equation}
since $\mathbb{Q}$ is dense in $\mathbb{R}$. Finally, writing the inequality
\begin{multline}
        \mathbf{\hat{x}}^{(k+1)}= \mathrm{sat}\left(\mathbf{A} \left( \mathbf{\hat{x}}^{(k)}-\mathbf{1}\right)+2\cdot\mathbf{1}+\mathbf{R}\mathbf{\hat{\boldsymbol\lambda}}\right)\\\leq  \mathrm{sat}\left(\mathbf{A} \left( \mathbf{{x}}^{(k)}-\mathbf{1}\right)+2\cdot\mathbf{1}+\mathbf{R}\mathbf{{\boldsymbol\lambda}}\right)=\mathbf{{x}}^{(k+1)}
\end{multline}
and recalling that $\mathbf{\hat{x}}^{(k)}$ achieves $\mathbf{1}$ in at most
$K=(2Q)^N$ iterations, we conclude that also $\mathbf{{x}}^{(k)}$ must achieve
$\mathbf{1}$ in at most $K$ iterations, by the Squeeze Theorem applied to
$\mathbf{\hat{x}}^{(k)} \leq \mathbf{{x}}^{(k)}\leq \mathbf{1}$. \end{IEEEproof}

Theorem \ref{thm - chaotic behaviors} states that there cannot exist bad equilibria, limit cycles or
chaotic behaviors (in short, \emph{bad trajectories}) if the
minimum channel LLR exceeds the solution $\tau$ of Problem
\ref{prb - original problem} or \ref{prb - generalized problem}.
This reinforces the name \emph{threshold} assigned to $\tau$ (we do
not distinguish any more between $\tau'$ and $\tau^*$), that is not limited to equilibria, but  pertains
to all bad trajectories.

\begin{figure}
\centering
  \includegraphics[width=\figwidthlarge]{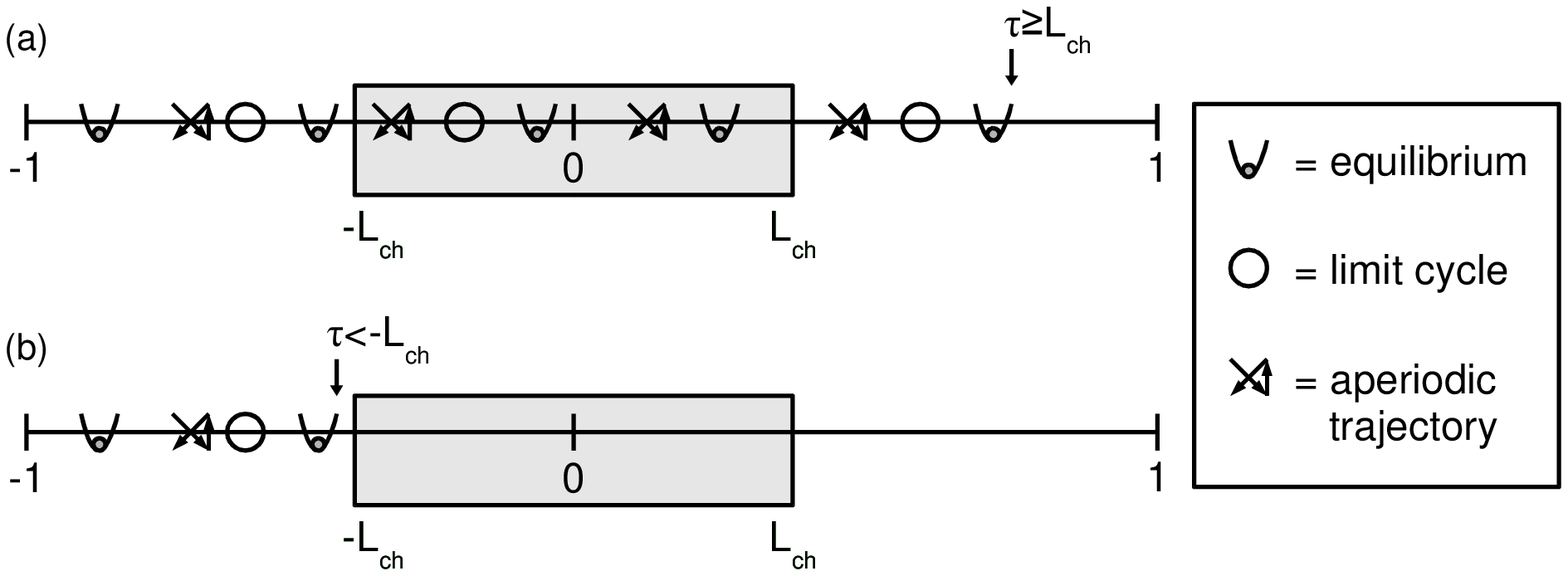}\\
  \caption{Trajectories of an AS, ordered w.r.t. their minimum channel LLR.}\label{fig - AS line}
\end{figure}

In Fig. \ref{fig - AS line}(a)  we represent bad trajectories, ordering them w.r.t. their minimum channel LLR, in the range
$-1\leq\lambda\leq 1$.
By Theorems \ref{thm - sequential cycles do not matter}
and \ref{thm - chaotic behaviors},
the rightmost bad trajectory is
 an equilibrium.

The results found so far can be exploited to deactivate many ASs
during the decoding process, using two different saturation
levels. Without loss of generality we set the saturation level of
extrinsic messages equal to $\pm 1$, and the saturation level of
channel LLRs equal to $\pm L_{ch}$, with $0<L_{ch}\leq 1$.
The latter saturation
level defines the range of admissible channel LLRs, depicted in Figs. \ref{fig - AS line}(a) and \ref{fig - AS line}(b) as a gray box. The decoding
trajectories within ASs can be very different in case of positive or negative
thresholds:
\begin{itemize}
  \item if $\tau\geq 0$, the saturation of channel LLRs to $\pm
      L_{ch}$ does not destroy bad trajectories. This is
      graphically represented in Fig. \ref{fig - AS line}(a);
  \item  if $\tau< 0$, we can set $L_{ch}<-\tau=|\tau|$. With this
      choice,  \emph{channel LLRs can never
lead to bad trajectories}, as depicted in Fig.
      \ref{fig - AS line}(b). Therefore, by Theorem \ref{thm - chaotic
      behaviors} the AS is defused.
\end{itemize}

%%%%%%%%%%%%%%%%%%%%%%%%%%%%%%%%%%%%%%%%%%%%%%%%%%%%%%%%%%%%%%%%%%%%%%%%%%%%%%%%%%%%%%%%%%%

\section{Simulation Results} \label{sec - Examples}
The behavior of bad structures under iterative decoding in a
 large code graph is in good agreement with the theory developed so far. For instance, consider the AS
(5,3) with topology shown in Fig. \ref{fig - messages}.
\begin{figure}
\centering
  \includegraphics[width=\figwidthsmall,clip]{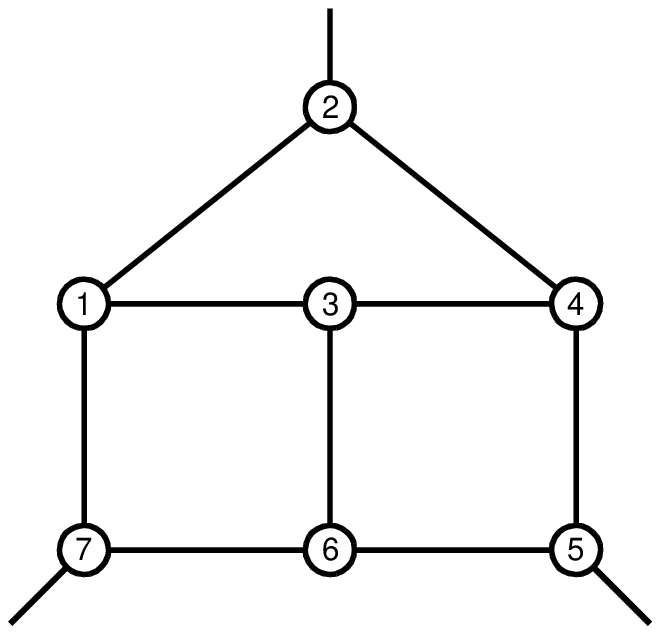}\\
  \caption{A (7,3) elementary absorbing set.}\label{fig - AS (7,3) topology}
\end{figure}
For this AS, $\tau = -1/3$ (a method to compute thresholds will be presented in the next Section).

\begin{figure}
\centering
  \includegraphics[width=\figwidthlarge,clip]{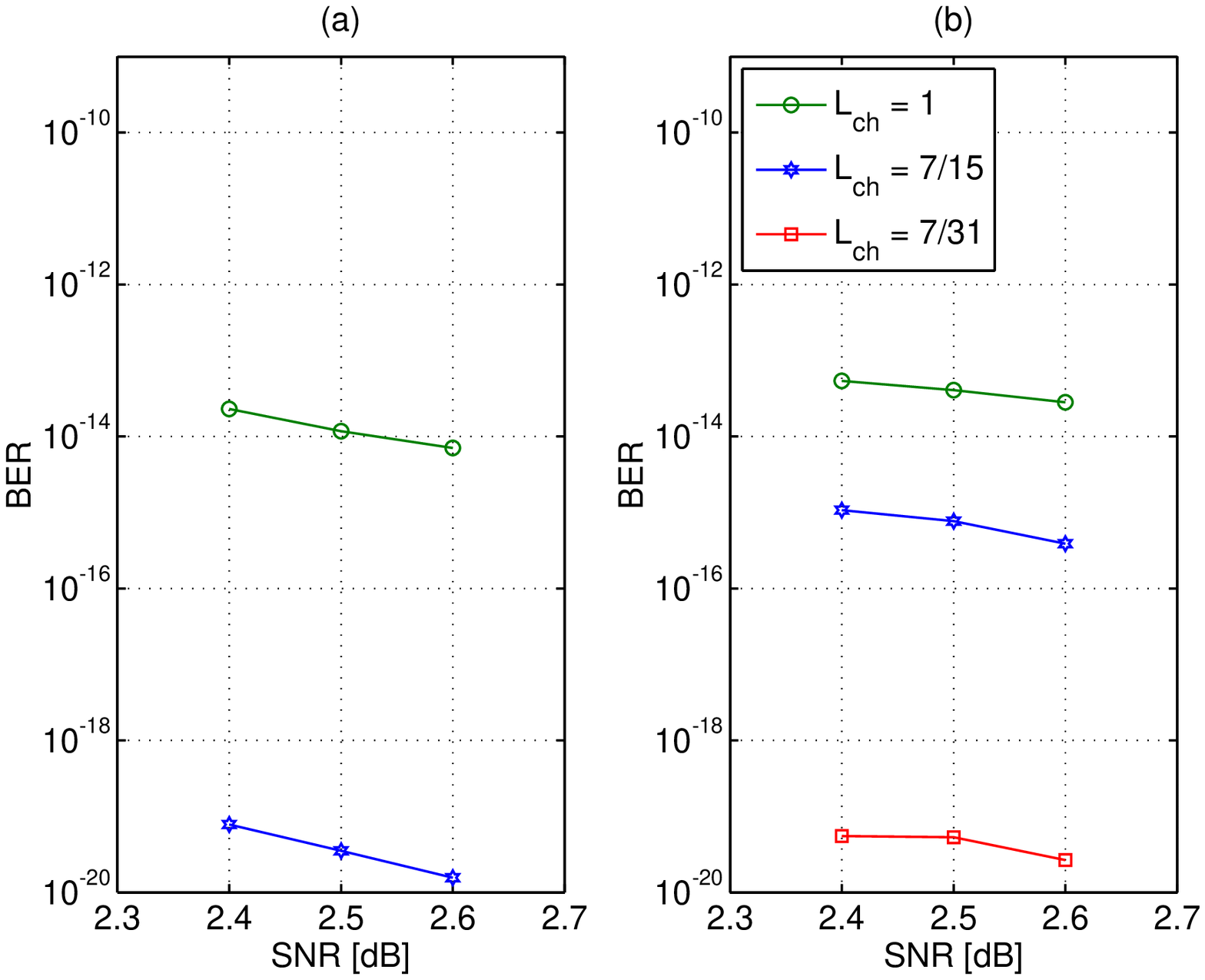}\\
  \caption{Error floor contributions of the (5,3) and (7,3) ASs  shown in Figs. \ref{fig - messages} and  \ref{fig - AS (7,3) topology}, respectively. The error floors have been obtained applying Importance Sampling to a real LDPC code, under MS sequential  decoding.}\label{fig - AS real behaviour}
\end{figure}

In Fig. \ref{fig - AS real behaviour}(a) we plot its contribution to the
error-floor of an LDPC code having block size $30\,000$ and rate $4/5$.
The simulations are run using
Importance Sampling over a Gaussian channel, with SNR around 2.5
dB. We always let the quantized channel LLRs vary in the range $\left[-7,7\right]$, while the extrinsic LLRs are quantized with a varying number  $q_e$ of bits. Therefore extrinsic messages belong to the interval $\left[-2^{q_e-1}+1,2^{q_e-1}-1\right]$, and $L_{ch}=\frac{7}{2^{q_e-1}-1}$.
Decisions are taken
after 20 iterations of MS sequential decoding.

From Fig. \ref{fig - AS real behaviour}(a) it is apparent that the probability that the MS
decoder be locked by the AS is lowered when $L_{ch}$ is reduced from $1$ to $7/15$. However, this is  larger
than $|\tau|=1/3$ and an error floor still appears. In agreement with the predictions of our
theory, if we set  $L_{ch}=7/31<|\tau|$
the AS is always unlocked and the error probability is
zero.

In Fig. \ref{fig - AS real behaviour}(b) we plot the
same curves for another AS embedded in the same LDPC code. This AS is shown in Fig. \ref{fig - AS (7,3) topology} and its  $\tau=-1/9$. Once again, reducing  $L_{ch}$ decreases the error probability, but now $L_{ch}=7/31$ does not guarantee an error-free performance.

\begin{figure}
\centering
  \includegraphics[width=\figwidthlarge,clip]{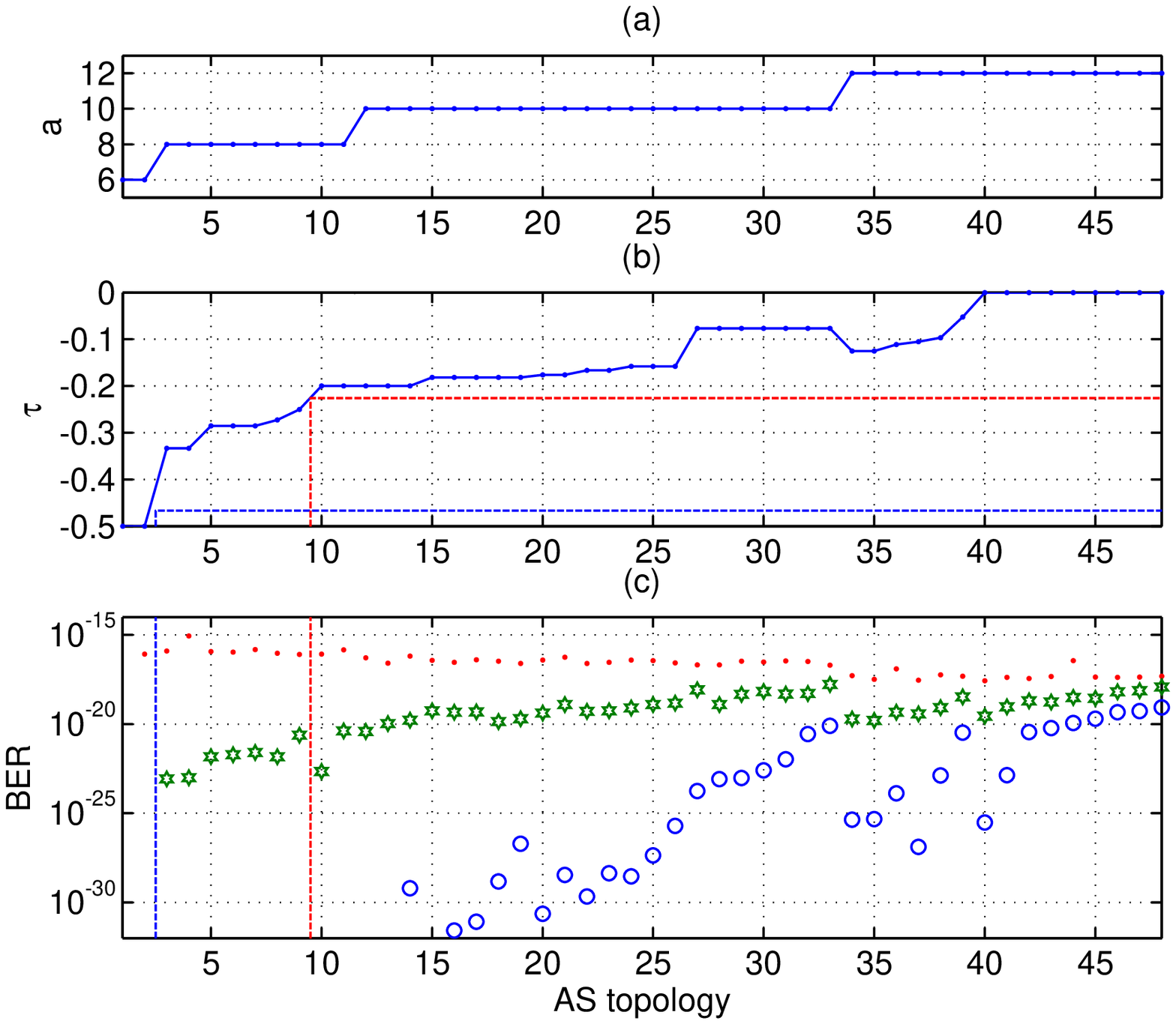}
  \caption{Error floor contributions of 48 ASs, with various channel saturation levels.}\label{fig - ASs contribution}
\end{figure}

Fig. \ref{fig - ASs contribution} refers to a different code,
with the same blocklength and rate. Here we have 48
different AS topologies of size $a=6,8,10,12$ (see Fig. \ref{fig - ASs contribution}(a)), $b=4$, whose thresholds are shown in
Fig. \ref{fig - ASs contribution}(b). In Fig. \ref{fig - ASs contribution}(c) we plot the BER contribution of each topology, with various channel saturation levels.  The results agree with the predictions of our model. If $L_{ch}=1$, all ASs contribute to the error floor. If $L_{ch}=7/15$ all the (6,4) ASs, whose threshold $\tau=-1/2$, are deactivated. With $L_{ch}=7/31$ all (8,4) ASs below threshold are deactivated. Also some ASs with threshold just above $-7/31$ gave no errors. Besides, Fig. \ref{fig - ASs contribution}(c) shows a good correlation between the thresholds and the dangerousness of the ASs.

%%%%%%%%%%%%%%%%%%%%%%%%%%%%%%%%%%%%%%%%%%%%%%%%%%%%%%%%%%%%%%%%%%%%%%%%%%%%%%%%%%%%%%%%%%%

\section{A search algorithm for thresholds}\label{sec - Search algorithm}

\subsection{Towards an affordable linear problem}
With the aim of deriving an efficient algorithm to compute the AS threshold,
we further simplify Problem \ref{prb - generalized problem}, introducing
\begin{Problem}\label{prb - inequality generalized problem}
    \begin{align}\nonumber
    \tilde{\tau}=    \max_{\lambda,\mathbf{x}}&\quad\lambda\\ s.t.
        & \quad -\mathbf{1}\leq\mathbf{x}\leq \mathbf{1} \label{eq - x bounds}\\
&\quad \exists j: x_j< 1 \label{eq - x < 1} \\
                    & \quad \mathbf{x}\geq\mathrm{\widetilde{sat}}\left(\mathbf{A} \left( \mathbf{x}-\mathbf{1}\right)+(2+\lambda)\mathbf{1}\right)\label{eq - inequality sat tilde}
 \end{align}\end{Problem}
\noindent where $\mathrm{\widetilde{sat}}(x)\triangleq\min(x,1)$.
With respect to Problem \ref{prb - generalized problem},  only the upper saturation is still present in $
\mathrm{\widetilde{sat}}(\cdot)$: extrinsic
messages can now assume any negative value. Besides, the
constraint imposed by the equilibrium equality has been relaxed,
and substituted by an inequality containing only a scalar
value $\lambda$. Notwithstanding these modifications, the following
theorems hold:
\begin{Theorem}\label{thm - tau*=tau tilde}
As for Problems \ref{prb - generalized problem} and \ref{prb -
inequality generalized problem}, $\tau^*= \tilde{\tau}$.
\end{Theorem}
\begin{IEEEproof}We show that $\tau^*\leq\tilde{\tau}$ and  $\tau^*\geq\tilde{\tau}$.

Assume we are given a solution $(\mathbf{x}^*,\boldsymbol\lambda^*)$ of Problem \ref{prb - generalized problem}, i.e. with $\min(\boldsymbol\lambda^*)=\tau^*$. We can exhibit a pair $(\mathbf{\tilde{x}},\tilde{\lambda})$ that satisfies the constraints of Problem \ref{prb - inequality generalized problem}. Indeed
\begin{multline}
  \mathbf{x}^*=\mathrm{sat}\left(\mathbf{A} \left( \mathbf{x}§^*-\mathbf{1}\right)+2\cdot\mathbf{1}+\boldsymbol\lambda^*\right)\\\geq\mathrm{sat}\left(\mathbf{A} \left( \mathbf{x}^*-\mathbf{1}\right)+\left(2+\min(\boldsymbol\lambda^*)\right)\mathbf{1}\right)\\\geq\mathrm{\widetilde{sat}}\left(\mathbf{A} \left( \mathbf{x}^*-\mathbf{1}\right)+\left(2+\min(\boldsymbol\lambda^*)\right)\mathbf{1}\right)\,.
\end{multline}
Therefore, the pair $\left(\mathbf{\tilde{x}}=\mathbf{x}^*,\tilde{\lambda}=\min(\boldsymbol\lambda^*)\right)$ fulfills the constraints of Problem  \ref{prb - inequality generalized problem}, because also $\exists j: \tilde{x}_j=x_j^*< 1$. Being $ \tilde{\tau}$ the result of a maximization, $\tilde{\tau}\geq \tilde{\lambda}= \min(\boldsymbol\lambda^*)=\tau^*$.

Focusing on the converse, assume we are given a solution
$(\mathbf{\tilde{x}},\tilde{\lambda})$ of Problem \ref{prb - inequality
generalized problem}, with $\tilde{\lambda}=\tilde{\tau}$. No matter whether
extrinsic messages are saturated or not, we can always add a positive vector
$\mathbf{\Delta}\geq\mathbf{0}$ to $\tilde{\lambda}\cdot\mathbf{1}$:
\begin{equation}
    \mathbf{\Delta}=\mathbf{\tilde{x}}-\mathrm{\widetilde{sat}}\left(\mathbf{A} \left( \mathbf{\tilde{x}}-\mathbf{1}\right)+(2+\tilde{\lambda})\mathbf{1}\right)\,.
\end{equation}
This way, we turn  inequality \eqref{eq - inequality sat tilde} into the equality
\begin{equation}
    \mathbf{\tilde{x}}=\mathrm{\widetilde{sat}}\left(\mathbf{A} \left( \mathbf{\tilde{x}}-\mathbf{1}\right)+2\cdot\mathbf{1}+\tilde{\lambda}\cdot\mathbf{1}+\mathbf{\Delta}\right)
\,.\end{equation}
The constraints of Problem \ref{prb - inequality
generalized problem} set $\mathbf{x}\geq -\mathbf{1}$, thus we conclude that
\begin{equation}
  \mathrm{\widetilde{sat}}\left(\mathbf{A} \left( \mathbf{\tilde{x}}-\mathbf{1}\right)+2\cdot\mathbf{1}+\tilde{\lambda}\cdot\mathbf{1}+\mathbf{\Delta}\right)\geq -\mathbf{1}
\end{equation}
and finally
\begin{multline}
  \mathrm{\widetilde{sat}}\left(\mathbf{A} \left( \mathbf{\tilde{x}}-\mathbf{1}\right)+2\cdot\mathbf{1}+\tilde{\lambda}\cdot\mathbf{1}+\mathbf{\Delta}\right)\\=  \mathrm{sat}\left(\mathbf{A} \left( \mathbf{\tilde{x}}-\mathbf{1}\right)+2\cdot\mathbf{1}+\tilde{\lambda}\cdot\mathbf{1}+\mathbf{\Delta}\right)\,.
\end{multline}

To conclude, if a solution $(\mathbf{\tilde{x}},\tilde{\lambda})$ of Problem
\ref{prb - inequality generalized problem} with $\tilde{\lambda}=\tilde{\tau}$
exists, we can exhibit a generalized equilibrium
$(\mathbf{x}^*=\mathbf{\tilde{x}},\boldsymbol\lambda^*=\tilde{\lambda}\cdot\mathbf{1}+\mathbf{\Delta})$ solution
of Problem \ref{prb - generalized problem}, with
$\mathbf{\Delta}\geq\mathbf{0}$. Being $\tau^*$ the result of a maximization,  we conclude that
$\tau^*\geq\min(\boldsymbol\lambda^*)=\min(\tilde{\lambda}\cdot\mathbf{1}+\mathbf{\Delta})\geq\tilde{\lambda}=\tilde{\tau}$.
\end{IEEEproof}

Once again, Problems \ref{prb - generalized problem} and \ref{prb - inequality generalized problem} are not equivalent, as the  solutions are different (the second one is not even  a  generalized equilibrium). Anyway, the two thresholds are the same.

Problem \ref{prb -
inequality generalized problem} is still non-linear and multimodal. Besides,
equations are still not differentiable. We further elaborate, rewriting
Problem \ref{prb - inequality generalized problem} in another form that does
not rely on the $\widetilde{\mathrm{sat}}(\cdot)$ function: we define a
partition of  $\{1,\ldots,N\}$ in the two subsets $\mathcal{S}^{unsat}$ and
$\mathcal{S}^{sat}$ of unsaturated and saturated messages, respectively\footnote{The adoption of both $\mathcal{S}^{unsat}$ and $\mathcal{S}^{sat}$ is
redundant, since $\mathcal{S}^{sat}=\{1,\ldots,N\} \backslash
\mathcal{S}^{unsat}$, but sometimes we use both of them for compactness.}.   We
also introduce a permutation matrix $\mathbf{\Pi}$, that reorganizes
extrinsic messages, putting the unsaturated ones on top:
\begin{equation}
    \begin{bmatrix}
     \mathbf{x}_{\mathcal{S}^{unsat}} \\
      \mathbf{1} \\
    \end{bmatrix}
    =
    \mathbf{\Pi}\mathbf{x}\,.
\end{equation}
Accordingly, we  permute the routing matrix $\mathbf{A}$, and divide it in
four submatrices, having inputs/outputs saturated or not:
\begin{equation}
    \mathbf{\Pi}\mathbf{A}\mathbf{\Pi}^{-1}=\begin{bmatrix}
                 \mathbf{A}_{\mathcal{S}^{unsat},\mathcal{S}^{unsat}} & \mathbf{A}_{\mathcal{S}^{unsat},\mathcal{S}^{sat}} \\
                 \mathbf{A}_{\mathcal{S}^{sat},\mathcal{S}^{unsat}} & \mathbf{A}_{\mathcal{S}^{sat},\mathcal{S}^{sat}} \\
               \end{bmatrix}\,.
\end{equation}

We are now ready to introduce
\begin{Problem}\label{prb - linearized generalized problem}
    \begin{align}
 \nonumber    \dot{\tau}&= \max_{\substack{\mathcal{S}^{unsat}\subseteq \{1,\ldots,N\} \\\mathcal{S}^{unsat}\neq\emptyset}} \max_{\lambda,\mathbf{x}_{\mathcal{S}^{unsat}}}\quad\lambda\\
        s.t.
        &\quad  -\mathbf{1}\leq\mathbf{x}_{\mathcal{S}^{unsat}}\leq \mathbf{1} \label{eq - x unsat bound}\\
         & \quad (\mathbf{A}_{\mathcal{S}^{unsat},\mathcal{S}^{unsat}}-\mathbf{I})\left( \mathbf{x}_{\mathcal{S}^{unsat}}-\mathbf{1}\right)+(1+\lambda)\mathbf{1}\leq \mathbf{0}\label{eq - linear prb unsat constraint}    \\
        & \quad \mathbf{A}_{\mathcal{S}^{sat},\mathcal{S}^{unsat}} \left( \mathbf{x}_{\mathcal{S}^{unsat}}-\mathbf{1}\right)+(1+\lambda)\mathbf{1}\geq \mathbf{0}\,.\label{eq - linear prb sat constraint}
 \end{align}
\end{Problem}
Note that the use of $\mathcal{S}^{unsat}$ in the above
problem is slightly misleading: even if the outer (leftmost) maximization sets
$j\in \mathcal{S}^{unsat}$,  thanks to \eqref{eq - x unsat bound} the inner
maximization could achieve its maximum even in $x_j=1$, and not in $x_j<1$. We shall show that this relaxation does not impair the threshold
computation.

The following  theorem holds.
\begin{Theorem}\label{thm - alpha_tilde <= alpha_dot}
As for Problems \ref{prb - inequality generalized problem} and
\ref{prb - linearized generalized problem}, $\tilde{\tau}= \dot{\tau}$.
\end{Theorem}

\begin{IEEEproof}
 We only give a sketch of the proof, since it is simple but quite long.
 We show that  Problem \ref{prb - inequality generalized problem} implies
Problem \ref{prb - linearized generalized problem}, and vice-versa.

First, \eqref{eq - x < 1} means that $\mathcal{S}^{unsat}\neq\emptyset$.
Rewriting \eqref{eq - inequality sat tilde} in the modified order, we obtain
$     \mathbf{x}_{\mathcal{S}^{unsat}}\geq
                 \mathbf{A}_{\mathcal{S}^{unsat},\mathcal{S}^{unsat}}
\left( \mathbf{x}_{\mathcal{S}^{unsat}}-\mathbf{1}\right)+(2+\lambda)\mathbf{1}
$,
since in the first block of inequalities the $\mathrm{\widetilde{sat}}(\cdot)$
operator is useless. Therefore  \eqref{eq - linear prb unsat constraint} must be true. As for the second block of inequalities, they hold only if the argument of the
$\mathrm{\widetilde{sat}}(\cdot)$ exceeds $\mathbf{1}$, i.e.,
$                    \mathbf{A}_{\mathcal{S}^{sat},\mathcal{S}^{unsat}}
\left( \mathbf{x}_{\mathcal{S}^{unsat}}-\mathbf{1}\right)+(2+\lambda)\mathbf{1}\geq \mathbf{1}
$, that immediately leads to \eqref{eq - linear prb sat constraint}.

Analogous arguments hold for the converse. The only tricky point is the following. Let
$(\mathbf{\dot{x}},\dot{\lambda})$ be a maximizer for  Problem \ref{prb - linearized
generalized problem}, with $\dot{\lambda}=\dot{\tau}$, and $\mathcal{\dot{S}}^{unsat}$  the set corresponding to this solution. As already highlighted, $\mathcal{\dot{S}}^{unsat}$ could contain
indices referring to saturated variables. If at least one element of
$\mathbf{x}_{\mathcal{\dot{S}}^{unsat}}$ is not saturated, this
 does not impair the outer maximization, since the same solution
$(\mathbf{\dot{x}},\dot{\lambda})$ is a maximizer  with another pattern of
saturations $\mathcal{S}^{unsat}=\left\{j\in \mathcal{\dot{S}}^{unsat}:
\dot{x}_j<1\right\} \neq \emptyset$. In this case,  \eqref{eq - x < 1} is
satisfied. On the contrary, if \emph{all}
messages in $\dot{S}^{unsat}$ were saturated, we could achieve the maximum
$\lambda$  not respecting  \eqref{eq - x < 1},  obtaining
$\dot{\tau}\geq\tilde{\tau}$. We now prove that this cannot happen. Indeed, if
we set $\mathbf{x}_{\mathcal{S}^{unsat}}=\mathbf{1}$,   \eqref{eq - linear prb
unsat constraint} becomes  $\lambda\leq-1$. Yet, similarly to Theorem \ref{thm
- equilibrium -1}, there are always other legitimate solutions having
$\lambda=-1$ that do not violate the constraints of Problem \ref{prb -
linearized generalized problem},  e.g.
$\left(\mathbf{x}=-\mathbf{1},\lambda=-1\right)$, for which  \eqref{eq -
linear prb sat constraint}  has no meaning and \eqref{eq - linear prb unsat
constraint}  becomes $-2(\mathbf{A}-\mathbf{I})\mathbf{1}\leq \mathbf{0}$,
that is true because
$\mathbf{A}\mathbf{1}\geq\mathbf{1}$.  We conclude that
$\dot{\tau}\geq -1$ and that substituting $\mathbf{x}_{\mathcal{S}^{unsat}}<
\mathbf{1}$ with $\mathbf{x}_{\mathcal{S}^{unsat}}\leq \mathbf{1}$ does not
harm the threshold computation.
\end{IEEEproof}

Once again, we do
not distinguish any more between $\tau'$, $\tau^*$, $\tilde{\tau}$ and $\dot{\tau}$ since they match, and simply use $\tau$.

In principle, we could solve the inner maximization of Problem
\ref{prb - linearized generalized problem}, repeatedly running an optimization
algorithm suited to linear equality and inequality constraints (e.g., the
simplex algorithm), and retaining only the largest value of $\tau$ among all
possible configurations of saturated messages. This is practically
unfeasible for two reasons. First, optimization algorithms are time-consuming
and we should resort to them with caution. Besides, the number of
configurations to test grows exponentially with $N$. Solving Problem \ref{prb
- linearized generalized problem} with a brute-force search becomes
unpracticable even for moderate values of $N$. In the following, we develop  methods to discard most
$\mathcal{S}^{unsat}$ configurations.

\subsection{Pruning tests}
Test 1 exploits the following two theorems:
\begin{Theorem}\label{thm - w0 sufficient condition}
If $\mathbf{A}_{\mathcal{S}^{unsat},\mathcal{S}^{unsat}}$ contains at least
one row with all-zero elements, there are no solutions satisfying the
constrains of Problem \ref{prb - linearized generalized problem}.
\end{Theorem}
\begin{IEEEproof} The proof is a \emph{reductio ad absurdum}. Consider any row
of $\mathbf{A}_{\mathcal{S}^{unsat},\mathcal{S}^{unsat}}$ with null weight,
say the one corresponding to the $j$-th in $\mathbf{x}$. Then, by \eqref{eq -
linear prb unsat constraint},
\begin{equation}
    x_j\geq 2+\lambda\geq 1
\end{equation}
where the second inequality holds since $\lambda\geq-1$. The above result
$x_j\geq 1$ contradicts the hypothesis  $j\in \mathcal{S}^{unsat}$ . \end{IEEEproof}

\begin{Theorem}\label{thm - w1 sufficient condition}
If $\mathbf{A}_{\mathcal{S}^{unsat},\mathcal{S}^{unsat}}$ contains at least
one row with exactly one element equal to $1$, say $A_{j,h}$,  and if the
column vector $\mathbf{A}_{\mathcal{S}^{sat},h}$ has weight larger than $0$,
then there are no solutions satisfying the constraints of Problem \ref{prb -
linearized generalized problem}.
\end{Theorem}
\begin{IEEEproof} The proof is  a \emph{reductio ad absurdum}. Consider any non
null element of     $\mathbf{A}_{\mathcal{S}^{sat},h}$, say $A_{i,h}$. Note
that  the maximum weight of any row and column of $\mathbf{A}$ is 2, being the
VN degree $d=3$. Thus, either $A_{i,h}$ is the only non-null element of the
row $\mathbf{A}_{i,\mathcal{S}^{unsat}}$, or at most another element
$A_{i,k}=1$ exists in $\mathbf{A}_{i,\mathcal{S}^{unsat}}$. If
$\mathbf{A}_{i,\mathcal{S}^{unsat}}$ has weight 1, by \eqref{eq - linear prb
sat constraint}
\begin{equation}
    1\leq x_h-1+2+\lambda =x_h+1+\lambda\,.
\end{equation}
If $\mathbf{A}_{i,\mathcal{S}^{unsat}}$ has weight 2, by \eqref{eq - linear
prb sat constraint}
\begin{equation}
    1\leq x_h-1+x_k-1+2+\lambda\leq x_h+1+\lambda
\end{equation}
where the second inequality holds since $x_k\leq 1$.
In either case,
\begin{equation}
    x_j\geq x_h-1+2+\lambda\geq  1
\end{equation}
where the first inequality comes from \eqref{eq - linear prb unsat
constraint}. The above result $x_j\geq 1$ contradicts the hypothesis  $j\in
\mathcal{S}^{unsat}$ . \end{IEEEproof}

Theorems \ref{thm - w0 sufficient condition} and \ref{thm - w1 sufficient
condition} suggest sufficient conditions to discard configurations of
$\mathcal{S}^{unsat}$. The advantage of Test 1 is simplicity. The weakness of Test 1 is that it does not
take advantage of  previous maximizations, with other configurations of
$\mathcal{S}^{unsat}$.

Test 2 exploits the threshold $\tau$ discovered up to that time.
It starts initializing lower and upper bounds
($\mathbf{l}$ and $\mathbf{u}$, respectively) for the  minimum channel LLR
$\lambda$ and for $\mathbf{x}_{\mathcal{S}^{unsat}}$:
\begin{equation}
  \mathbf{l}\leq\begin{bmatrix}
                \lambda \\
                \mathbf{x}_{\mathcal{S}^{unsat}} \\
              \end{bmatrix}\leq\mathbf{u}
\end{equation}
where
\begin{equation}\label{eq - test 2 initialization}
  \mathbf{l}=\begin{bmatrix}
              \tau \\ -\mathbf{1}_{M \times 1}
            \end{bmatrix},\quad\quad\quad \mathbf{u}=\begin{bmatrix}
              \lambda_{max} \\ \mathbf{1}_{M \times 1}
            \end{bmatrix}
\end{equation}
and $M=\mathrm{card}\left(\mathcal{S}^{unsat}\right)$.
In the most general case, $\lambda_{max}=1$. Yet we are mainly interested in
the negative semi-axis, since in Section \ref{sec - Chaotic} we have
shown that we can deactivate an AS only if the threshold is  negative.
Therefore, in the  threshold computation algorithm we can lower
$\lambda_{max}$ (of course, we must keep $\lambda_{max}\geq 0$), exchanging
some information loss (we return $\min(\tau,\lambda_{max})=\lambda_{max}$ when $\tau>\lambda_{max}$), for an increased capability to
discard saturation patterns, resulting in an execution speedup.

Test 2 analyzes the inequality constraints of Problem \ref{prb - linearized generalized problem} in turn,
rewritten as
 \begin{equation}
    \mathbf{C}\begin{bmatrix}
                \lambda \\
                \mathbf{x}_{\mathcal{S}^{unsat}} \\
              \end{bmatrix}
               \leq
\mathbf{f}
 \end{equation}
 with
\begin{eqnarray}
   \mathbf{C}&=&
    \begin{bmatrix}

         \mathbf{1}_{M\times 1} & \mathbf{A}_{\mathcal{S}^{unsat},\mathcal{S}^{unsat}}-\mathbf{I} \\
         -\mathbf{1}_{(N-M)\times 1}  & -\mathbf{A}_{\mathcal{S}^{sat},\mathcal{S}^{unsat}} \\
    \end{bmatrix}\\
        \mathbf{f}&=&\begin{bmatrix}
                  \mathbf{A}_{\mathcal{S}^{unsat},\mathcal{S}^{unsat}}\mathbf{1}_{M\times 1}-2\cdot\mathbf{1}_{M\times 1} \\ \mathbf{1}_{(N-M)\times 1}-\mathbf{A}_{\mathcal{S}^{sat},\mathcal{S}^{unsat}}\mathbf{1}_{M\times 1} \\
                \end{bmatrix}\,.
 \end{eqnarray}

For every variable involved, the test tries to tighten the gap between the corresponding lower and upper
bounds, exploiting  bounds (upper or lower, depending on the coefficient signs)
on the other variables. The process can terminate in two ways:
\begin{enumerate}
  \item  bounds $\mathbf{l}$ and $\mathbf{u}$ cannot be further improved,
      and $\mathbf{l}\leq \mathbf{u}$:   $\tau$ and $\mathcal{S}^{unsat}$
      are compatible with the existence of other  equilibria,
      having thresholds larger than $\tau$;
  \item for at least one index $j$, we achieve $u_j<l_j$:  equilibria  having thresholds larger than the currently discovered $\tau$ cannot exist, for that $\mathcal{S}^{unsat}$.
\end{enumerate}
The initialization in \eqref{eq - test 2 initialization} influences
the algorithm effectiveness: the more the discovered threshold $\tau$   gets
large, the more Test 2 will effectively detect
impossible configurations of $\mathcal{S}^{unsat}$, speeding up the solution
of Problem \ref{prb - linearized generalized problem}.

\subsection{Tree based, efficient search of the AS threshold}
Test 2 is typically more effective than Test 1, as it can
detect a large number of configurations of $\mathcal{S}^{unsat}$  not
improving the threshold $\tau$. Yet, it can be applied only when
$\mathcal{S}^{unsat}$ is formed. On the contrary, Test 1 can also be applied
during the construction of $\mathcal{S}^{unsat}$:
 \begin{Theorem}
Let $\mathcal{S}^{violation}\subseteq\mathcal{S}^{unsat}$ be the set of indices
satisfying Theorems \ref{thm - w0 sufficient condition} or  \ref{thm - w1
sufficient condition}. Erasing only a subset of $\mathcal{S}^{violation}$ from
$\mathcal{S}^{unsat}$, the other elements in $\mathcal{S}^{violation}$ still
satisfy conditions of Theorems \ref{thm - w0 sufficient condition} or
\ref{thm - w1 sufficient condition}.
 \end{Theorem}
 \begin{IEEEproof} Assume that only one element involved in some violation, say
the $m$-th, is erased by $\mathcal{S}^{unsat}$. The proof for a generic subset
of violations, erased all together, can be achieved repeating the following
argument many times, discarding one element after the other.

When element $m$ passes from $\mathcal{S}^{unsat}$ to $\mathcal{S}^{sat}$, not
only the row $\mathbf{A}_{m,\mathcal{S}^{unsat}}$ must be erased, but also the
column $\mathbf{A}_{\mathcal{S}^{unsat},m}$ must be canceled. Looking at any
other row of  $\mathbf{A}_{\mathcal{S}^{unsat},\mathcal{S}^{unsat}}$ leading
to a violation, say the one corresponding to element $j$, three events can
happen (see Fig. \ref{fig - row column deletion}):
\begin{figure}
  \centering
  \includegraphics[width=\figwidthlarge]{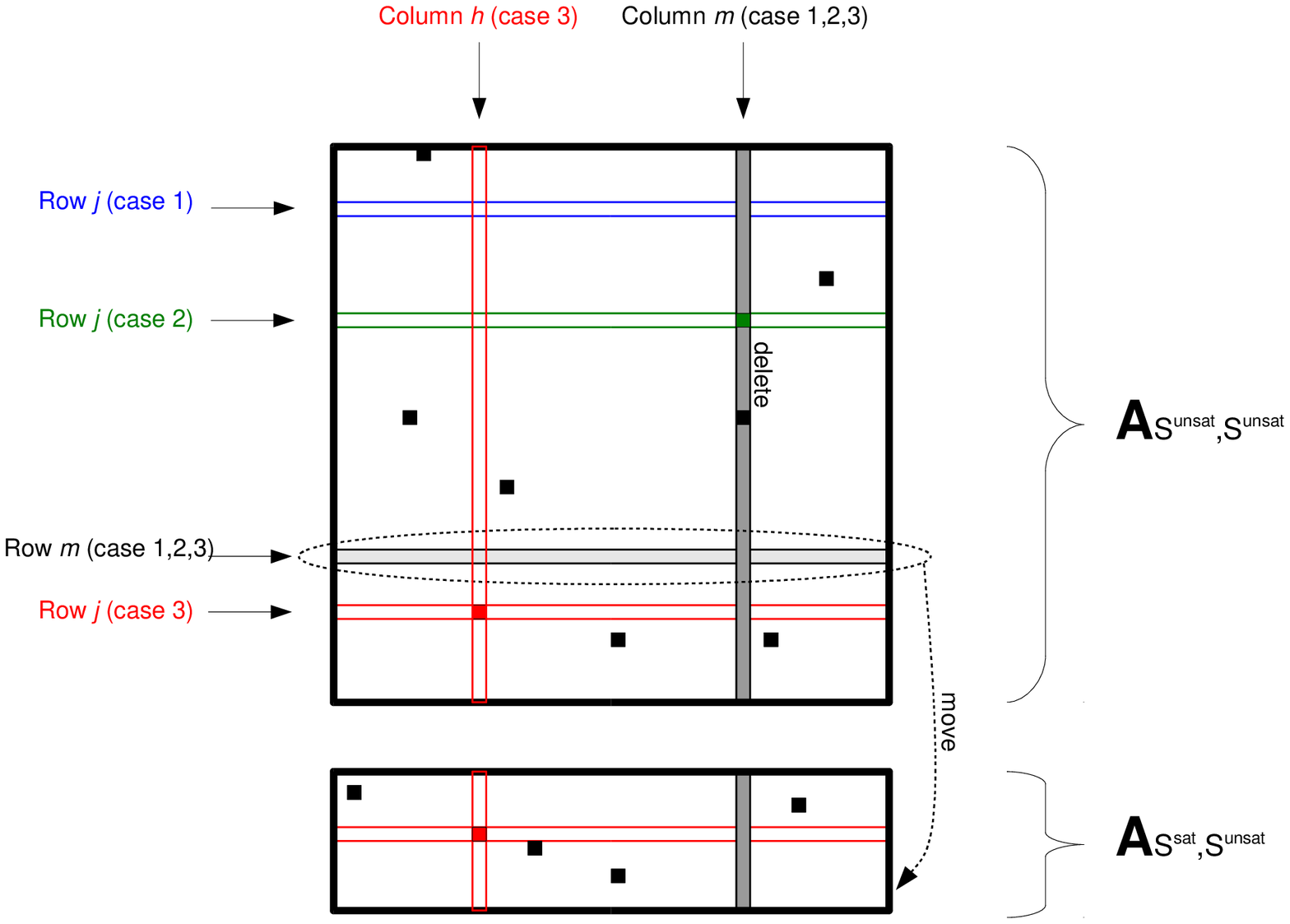}\\
  \caption{Row-column erasure possibilities in $\mathbf{A}_{\mathcal{S}^{unsat},\mathcal{S}^{unsat}}$.}\label{fig - row column deletion}
\end{figure}
\begin{enumerate}
  \item $\mathbf{A}_{j,\mathcal{S}^{unsat}}$ had weight 0: after column deletion, it still has weight 0  and the hypothesis of Theorem \ref{thm - w0 sufficient condition} is still valid;
  \item  $\mathbf{A}_{j,\mathcal{S}^{unsat}}$ had weight 1, and its element equal to 1 was exactly in the $m$-th column (the erased one): after deletion, the row assumes weight 0, therefore satisfying the hypothesis of Theorem \ref{thm - w0 sufficient condition};
  \item $\mathbf{A}_{j,\mathcal{S}^{unsat}}$ had weight 1, and the element equal to 1, say $A_{j,h}$ did not lie in the $m$-th column: after deletion, the row of $\mathbf{A}_{\mathcal{S}^{unsat},\mathcal{S}^{unsat}}$  still has weight 1. Since the weight of the corresponding  column  $\mathbf{A}_{\mathcal{S}^{sat},h}$ is still 1 (it had weight 1 by hypothesis), Theorem \ref{thm - w1 sufficient condition}  still holds for the $j$-th element.
\end{enumerate}
Either Theorem \ref{thm - w0 sufficient condition} or
\ref{thm - w1 sufficient condition} are still valid, and the other violations
do not disappear.
 \end{IEEEproof}
The above Theorem gives us the freedom to erase  elements in
$\mathcal{S}^{violation}$ all together from  $\mathcal{S}^{unsat}$, and
simultaneously add them to  $\mathcal{S}^{sat}$. Therefore, we can imagine a
\emph{tree search} among all possible configurations of saturated messages.

At the root node, $\mathcal{S}^{sat}=\emptyset$. At successive steps, some
extrinsic messages are marked as already visited (``fixed", from here on). In
addition, fixed messages are labeled as saturated or not. Extrinsic messages
not fixed (say ``free") are always unsaturated. This implicitly defines
 $\mathcal{S}^{sat}$ and $\mathcal{S}^{unsat}$.
For the current configuration, Test 1 is performed. Three things can happen:
\begin{itemize}
  \item \textbf{Case 1}: Test 1  claims that Problem \ref{prb - linearized generalized problem} may have solutions for that  $\mathcal{S}^{sat}$;
  \item \textbf{Case 2}: Test 1 claims that $\mathcal{S}^{sat}$ is incompatible with any solution of Problem \ref{prb - linearized generalized problem}, and all the elements generating a test violation are free;
  \item \textbf{Case 3}:  Test 1 claims that $\mathcal{S}^{sat}$ is incompatible with any solution of Problem \ref{prb - linearized generalized problem}, but some elements generating a test violation have been previously fixed (and marked as unsaturated).
\end{itemize}
Depending on the answer of Test 1, we expand the tree in different manners:
\begin{itemize}
  \item in Case 1, in turn we fix one of the free messages, and branch the tree, labeling the last element as either saturated or not, calling the algorithm recursively;
  \item in Case 2, we fix and mark as saturated all elements of  $\mathcal{S}^{unsat}$ that generate violations, and call the algorithm recursively;
  \item in Case 3, or if all variables have been already fixed, we take no action.
\end{itemize}
After Test 1, before the tree branching, we either perform optimization or not:
\begin{itemize}
  \item in Case 1 or 2, Test 2 is executed. In case of negative result, we return $\tau=-1$; otherwise, the simplex algorithm is eventually performed to solve Problem \ref{prb - linearized generalized problem} for the current $\mathcal{S}^{unsat}$;
  \item in Case 3, we return the partial result $\tau=-1$.
\end{itemize}

This way, Test 1 speeds up the
construction of $\mathcal{S}^{sat}$ and prunes many branches. Test 2 avoids the
execution of the simplex algorithm for many useless configurations,
not detected by Test 1.

%%%%%%%%%%%%%%%%%%%%%%%%%%%%%%%%%%%%%%%%%%%%%%%%%%%%%%%%%%%%%%%%%%%%%%%%%%%%%%%%%%%%%%%%%%%

\section{Additional properties}\label{sec - Other Properties}
\subsection{Punctured LDPC codes}

Puncturing is a popular means to
adapt the code rate or even achieve rate compatibility
\cite{Bib-PisFekTIT07}. An interesting extension of our theory is
that harmless ASs having $\tau<-L_{ch}$ are deactivated even in case of puncturing.

\begin{Theorem}\label{thm - puncturing}
Puncturing at most $a-1$ VNs of an AS does not increase the threshold.
\end{Theorem}
\begin{IEEEproof}
%The proof is a \emph{reductio ad absurdum}.
Assume that an AS $(a,b)$ of
threshold $\tau$ is punctured in less than $a$ VNs. Let
$\boldsymbol\lambda^{p}$ be the set of channel LLRs,
with null messages for the punctured VNs. First, consider the
case $\tau<0$. Assume that after puncturing, a bad trajectory
$\{\mathbf{x}^{(k)}\}$ exists, with
$\min(\boldsymbol\lambda^{p})>\tau$. This is an absurdum,
since $\boldsymbol\lambda^{p}$ is  a legitimate
solution without puncturing, and the definition
of threshold given e.g. in Problem \ref{prb - original problem} is
contradicted. This holds as long as at least one variable is not
punctured, otherwise nothing is left to optimize and
$\boldsymbol\lambda^{p}=\mathbf{0}>\tau\cdot\mathbf{1}$.

Consider now the case $\tau\geq 0$. Since at least one entry of $\boldsymbol\lambda^{p}$ is equal to $0$, we have $\min(\boldsymbol\lambda^{p})\leq 0$. Thus $\min(\boldsymbol\lambda^{p})\leq \tau$ and the threshold of the same AS without puncturing is not exceeded.
\end{IEEEproof}
Therefore, ASs having
$\tau<-L_{ch}$ cannot become harmful.

\subsection{Thresholds are rational numbers}
A final, not trivial property of
thresholds is the following.
\begin{Theorem}\label{thm - rational thresholds}
Thresholds $\tau\in \mathbb{Q}$.
\end{Theorem}
\begin{IEEEproof}
We focus on Problem \ref{prb - linearized generalized problem}. We will prove the theorem for any constrained saturation pattern $S^{unsat}$. Therefore, the result will hold for the maximum across all possible $S^{unsat}$. The proof is slightly cumbersome, and involves standard concepts of linear programming theory.

  First, note that constraints $-\mathbf{1} \leq \mathbf{x}\leq \mathbf{1}$ bound the feasible space of extrinsic messages. By Theorem \ref{thm - alpha range}, the constraint $-1 \leq \lambda \leq 1$ can be added without modifying the result. Therefore, the above constraints and the others in Problem \ref{prb - linearized generalized problem} define a \emph{polytope} $\mathcal{P}$ in $M+1$ dimensions
\begin{equation}
  \mathbf{y}\in \mathcal{P}\Longleftrightarrow  \underbrace{\begin{bmatrix}
                                                   \mathbf{I}_{(M+1)\times (M+1)}  \\
                             -\mathbf{I}_{(M+1)\times (M+1)}   \\
                                                  \mathbf{C} \\
                                                \end{bmatrix}}_{\triangleq\mathbf{C}'}
   \mathbf{y}\leq \underbrace{\begin{bmatrix}
 \mathbf{1}_{(M+1)\times 1} \\
                            \mathbf{1}_{(M+1)\times 1} \\
                    \mathbf{f} \\
                  \end{bmatrix}}_{\triangleq\mathbf{f}'}
\end{equation}
where
\begin{equation}
  \mathbf{y} \triangleq \begin{bmatrix}
                          \lambda \\
                          \mathbf{x}_{\mathcal{S}^{unsat}} \\
                        \end{bmatrix}
\end{equation}
Geometrically, the above inequality constraints reported in canonical form represent half-spaces that ``shave" the polytope.
The polytope $\mathcal{P}$ is convex, since it results from the intersection of half-spaces, that are affine and therefore convex.
To conclude, our optimization problem can be re-stated as
\begin{equation}
   \max_{\mathbf{y}}\;\mathbf{w}^T\mathbf{y},\quad\quad
        s.t.
       \quad \mathbf{y}\in\mathcal{P}
\end{equation}
with $\mathbf{w}=\begin{bmatrix}
                   1 & 0 & \ldots & 0 \\
                 \end{bmatrix}^T
$. Our feasible region cannot be empty, since we already know that a solution $(\lambda=-1, \mathbf{x}_{\mathcal{S}^{unsat}}=-\mathbf{1})$, i.e. $\mathbf{y}=-\mathbf{1}$ always exists.

From Linear Programming Theory \cite{LinearProgramming}, we know that
the number of independent constraints at any vertex is $M+1$ and that
at least one vertex is an optimizer in linear programming problems (the latter part is the enunciation of the Fundamental Theorem of Linear Programming).

     Focus on a vertex $\mathbf{v}$ that is also a maximizer, and on the  $M+1$ linearly independent constraints satisfied with equality in that point. Let $\mathcal{A}$ be the set of these constraints. We can write $\mathbf{C}'_{\mathcal{A},\{1,\ldots,M+1\}}\mathbf{v}=\mathbf{f}'_{\mathcal{A}}$.     Being  $\mathbf{C}'_{\mathcal{A},\{1,\ldots,M+1\}}$ full-rank, we can achieve a full QR-like decomposition $\mathbf{C}'_{\mathcal{A},\{1,\ldots,M+1\}}=\mathbf{Q}\mathbf{L}$, being $\mathbf{Q}$ orthogonal, and $\mathbf{L}$ lower triangular. Note that, being  the entries of   $\mathbf{C}'_{\mathcal{A},\{1,\ldots,M+1\}}$ rational (actually, integer),  we can always keep the elements of $\mathbf{Q}$ and $\mathbf{L}$ rational, e.g. performing a Gram-Schmidt decomposition.      Multiplying both sides of the above equation by $\mathbf{Q}^T$, we obtain
      \begin{equation}
      \mathbf{L}  \mathbf{v}=\mathbf{Q}^T\mathbf{f}'_{\mathcal{A}}\,.
      \end{equation}
     Focusing on the first line of the above system,   we achieve  $v_1=\lambda \in \mathbb{Q}$, since also $\mathbf{f}'\in \mathbb{Q}$.
\end{IEEEproof}

%%%%%%%%%%%%%%%%%%%%%%%%%%%%%%%%%%%%%%%%%%%%%%%%%%%%%%%%%%%%%%%%%%%%%%%%%%%%%%%%%%%%%%%%%%%

\section{Conclusions} \label{sec - Conclusions}
In this paper we defined a simplified model
for the evolution inside an absorbing set of the messages of an LDPC Min-Sum decoder, when saturation is applied. Based on this model we identified a parameter for
each AS topology, namely a \emph{threshold}, that is the result of
a \emph{max-min} non linear problem, for which we proposed an
efficient search algorithm. We have shown that based on this threshold
it is possible to classify the AS dangerousness. If all the channel LLRs inside the AS are  above this threshold, after a sufficient number of iterations the MS decoder can not be trapped by the AS.

Future work will primarily focus on the
extension of these concepts to \emph{scaled} and \emph{offset} MS
decoders. We are also trying to further simplify the threshold evaluation.

%%%%%%%%%%%%%%%%%%%%%%%%%%%%%%%%%%%%%%%%%%%%%%%%%%%%%%%%%%%%%%%%%%%%%%%%%%%%%%%%%%%%%%%%%%%


\begin{thebibliography}{10}
\providecommand{\url}[1]{#1}
\csname url@samestyle\endcsname
\providecommand{\newblock}{\relax}
\providecommand{\bibinfo}[2]{#2}
\providecommand{\BIBentrySTDinterwordspacing}{\spaceskip=0pt\relax}
\providecommand{\BIBentryALTinterwordstretchfactor}{4}
\providecommand{\BIBentryALTinterwordspacing}{\spaceskip=\fontdimen2\font plus
\BIBentryALTinterwordstretchfactor\fontdimen3\font minus
  \fontdimen4\font\relax}
\providecommand{\BIBforeignlanguage}[2]{{%
\expandafter\ifx\csname l@#1\endcsname\relax
\typeout{** WARNING: IEEEtran.bst: No hyphenation pattern has been}%
\typeout{** loaded for the language `#1'. Using the pattern for}%
\typeout{** the default language instead.}%
\else
\language=\csname l@#1\endcsname
\fi
#2}}
\providecommand{\BIBdecl}{\relax}
\BIBdecl

\bibitem{Bib-FreKoetTIT01}
B.~J. Frey, R.~Koetter, and A.~Vardy, ``Signal-space characterization of
  iterative decoding,'' \emph{IEEE Trans. Inf. Theory}, vol.~47, no.~2, pp.
  766--781, Feb. 2001.

\bibitem{Bib-KoetVonISTC03}
R.~Koetter and P.~. Vontobel, ``Graph covers and iterative decoding of
  finite-length codes,'' in \emph{$3^{rd}$ International Symposium on Turbo
  Codes \& Related Topics}, Brest, France, Sept. 2003, pp. 75--82.

\bibitem{RichErrorFloors}
T.~Richardson, ``Error floors of {LDPC} codes,'' in \emph{$41^{st}$ Annual
  Allerton Conf. on Commun., Contr. and Computing,}, Oct. 2003, pp. 1426--1435.

\bibitem{Vasic06}
S.~Chilappagari, S.~Sankaranarayanan, and B.~Vasic, ``{Error Floors of {LDPC}
  Codes on the Binary Symmetric Channel},'' in \emph{IEEE International
  Conference on Communications ({ICC'06})}, Istanbul, Turkey, June 2006, pp.
  1089--1094.

\bibitem{ChiVasJSAC09}
S.~Chilappagari, M.~Chertkov, M.~Stepanov, and B.~Vasic, ``Instanton-based
  techniques for analysis and reduction of error floors of {LDPC} codes,''
  \emph{IEEE J. Select. Areas Commun.}, vol.~27, no.~6, pp. 855--865, June
  2009.

\bibitem{Dol_ICC07}
L.~Dolecek, Z.~Zhang, M.~Wainwright, V.~Anantharam, and B.~Nikolic, ``Analysis
  of absorbing sets for array-based {LDPC} codes,'' in \emph{IEEE International
  Conference on Communications ({ICC'07})}, Glasgow, Scotland, U.K., June 2007,
  pp. 6261--6268.

\bibitem{Dol_JSAC09}
L.~Dolecek, P.~Lee, Z.~Zhang, V.~Anantharam, B.~Nikolic, and M.~Wainwright,
  ``Predicting error floors of structured {LDPC} codes: deterministic bounds
  and estimates,'' \emph{IEEE J. Select. Areas Commun.}, vol.~27, no.~6, pp.
  908--917, Aug. 2009.

\bibitem{Dol_TCOM09}
Z.~Zhang, L.~Dolecek, B.~Nikolic, V.~Anantharam, and M.~Wainwright, ``Design of
  {LDPC} decoders for improved low error rate performance: quantization and
  algorithm choices,'' \emph{IEEE Trans. Wireless Commun.}, vol.~57, no.~11,
  pp. 3258--3268, Nov. 2009.

\bibitem{Schlegel}
C.~Schlegel, ``On the dynamics of the error floor behavior in (regular) {LDPC}
  codes,'' \emph{IEEE Trans. Inf. Theory}, vol.~56, no.~7, pp. 3248--3264, July
  2010.

\bibitem{Dolecek}
L.~Dolecek, Z.~Zhang, V.~Anantharam, M.~Wainwright, and B.~Nikolic, ``Analysis
  of absorbing sets and fully absorbing sets of array-based {LDPC} codes,''
  \emph{IEEE Trans. Inf. Theory}, vol.~56, no.~1, pp. 181--201, Jan. 2010.

\bibitem{Dolecek_PostProc}
Z.~Zhang, L.~Dolecek, B.~Nikolic, V.~Anantharam, and W.~M., ``Lowering {LDPC}
  error floors by postprocessing,'' in \emph{GLOBECOM 2008}, New Orleans, LA,
  Dec. 2008, pp. 1--6.

\bibitem{SunTahFitz}
J.~Sun, O.~Y. Takeshita, and M.~P. Fitz, ``Analysis of trapping sets for {LDPC}
  codes using a linear system model,'' in \emph{$42^{nd}$ Annual Allerton Conf.
  on Commun., Contr. and Computing,}, Oct. 2004.

\bibitem{Bib-ButSieALL11}
B.~Butler and P.~Siegel, ``Error floor approximation for {LDPC} codes in the
  {AWGN} channel,'' in \emph{$49^{th}$ Annual Allerton Conf. on Commun., Contr.
  and Computing,}, Oct. 2011, pp. 204--211.

\bibitem{Dol_ITA10}
L.~Dolecek, ``On absorbing sets of structured sparse graph codes,'' in
  \emph{Information Theory and Applications Workshop {(ITA)}}, UCSD, San Diego,
  CA, Feb. 1-5 2010, pp. 1--5.

\bibitem{McKPos}
D.~MacKay and M.~Postol, ``Weaknesses of {M}argulis and {R}amanujan-{M}argulis
  low-density parity-check codes,'' \emph{Electron. Notes in Theor. Comp.
  Sci.}, vol.~74, no.~10, pp. 97--104, Oct. 2003.

\bibitem{ShaLit}
E.~Sharon and J.~G. S.~Litsyn, ``Efficient serial {Message-Passing} schedules
  for {LDPC} decoding,'' \emph{IEEE Trans. Inf. Theory}, vol.~53, no.~11, pp.
  4076--4091, Nov. 2007.

\bibitem{Bib-PisFekTIT07}
H.~Pishro-Nik and F.~Fekri, ``Results on punctured {Low-Density Parity-Check}
  codes and improved iterative decoding techniques,'' \emph{IEEE Trans. Inf.
  Theory}, vol.~53, no.~2, pp. 599--614, Feb. 2007.

\bibitem{LinearProgramming}
K.~G. Murty, \emph{Linear programming}.\hskip 1em plus 0.5em minus 0.4em\relax
  New York: John Wiley \& Sons, 1983.

\end{thebibliography}
\end{document}